\documentclass[amsmath,amssymb,twocolumn]{aastex62} 
\usepackage{amsmath}
\usepackage{graphicx}
\usepackage{longtable}
\usepackage{bm}
\usepackage[caption=false,subrefformat=parens]{subfig}
\usepackage{hyperref}
\usepackage{nicefrac}
\usepackage{orcidlink}
\usepackage[normalem]{ulem}
\DeclareMathOperator{\arcsinh}{arcsinh}
\usepackage{lineno}
\journalinfo{ }
\hypersetup{colorlinks=true,breaklinks,linkcolor=red,citecolor=blue}

\begin{document}
  
\title{The thermal index of neutron-star matter in the virial approximation}
 
\author{Giuseppe Rivieccio \orcidlink{0009-0009-9456-6382}}
\affiliation{Departament d'Astronomia i Astrofísica, Universitat de València, C/ Dr Moliner 50, 46100, Burjassot (València), Spain}
\email{giuseppe.rivieccio@uv.es}

\author{Adriana Nadal-Matosas \orcidlink{0009-0007-5871-2587}}
\affiliation{ Departament de F\'isica Qu\`antica i Astrof\'isica (FQA), 
 Universitat de Barcelona (UB), 
 c. Mart\'i i Franqu\`es 1, E08028 Barcelona, Spain
}

\affiliation{Institut de Ci\`encies del Cosmos (ICCUB),
 Universitat de Barcelona (UB), 
 c. Mart\'i i Franqu\`es 1, E08028 Barcelona, Spain}
\author{Arnau Rios \orcidlink{0000-0002-8759-3202}}
\affiliation{ Departament de F\'isica Qu\`antica i Astrof\'isica (FQA), 
 Universitat de Barcelona (UB), 
 c. Mart\'i i Franqu\`es 1, E08028 Barcelona, Spain
}
\affiliation{Institut de Ci\`encies del Cosmos (ICCUB),
 Universitat de Barcelona (UB), 
 c. Mart\'i i Franqu\`es 1, E08028 Barcelona, Spain}
\email{arnau.rios@icc.ub.edu}
\author{Milton Ruiz
\orcidlink{0000-0002-7532-4144}}
\affiliation{Departament d'Astronomia i Astrofísica, Universitat de València, C/ Dr Moliner 50, 46100, Burjassot (València), Spain}

\begin{abstract}
Motivated by gravitational wave observations of binary neutron-star mergers, we study the thermal
index of low-density, high-temperature dense matter.  We use the virial expansion to account for nuclear interaction effects. We focus on the region of validity of the expansion, which reaches $10^{-3}$ fm$^{-3}$ at $T=5$ MeV up to almost saturation density at $T=50$ MeV. 
In pure neutron matter, we find an analytical expression for the thermal index, and show that 
it is nearly density- and temperature-independent,
within a fraction of a percent of the non-interacting, non-relativistic value of 
$\Gamma_\text{th} \approx 5/3$. 
When we incorporate protons, electrons and photons, we find that the density and temperature dependence of the thermal index changes significantly. 
We predict a smooth transition between an electron-dominated regime with
$\Gamma_\text{th} \approx 4/3$ at low densities to a neutron-dominated region with $\Gamma_\text{th} \approx 5/3$ at high densities. This behavior is by and large independent of proton fraction and is not affected by nuclear interactions in the region where the virial expansion converges. We model this smooth transition analytically and provide a simple but accurate parametrization of the inflection point between these regimes.  When compared to tabulated realistic models of the thermal index, we find an overall agreement at high temperatures that weakens for colder matter. The discrepancies can be attributed to the missing contributions
of nuclear clusters.  The virial approximation provides a clear and physically intuitive framework for understanding the thermal properties of dense matter, offering a computationally efficient solution that makes it particularly well-suited for the regimes relevant to neutron star binary remnants.
\end{abstract}
%

%
\section{Introduction}

The detection of the gravitational wave (GW) event GW170817\citep{LIGOScientific:2017vwq}, originating from a binary neutron star (BNS) merger and accompanied by electromagnetic (EM) counterpart radiation across the spectrum~\citep{LIGOScientific:2017zic,Coulter:2017wya,Savchenko:2017ffs,Lazzati:2016yxl,Murguia-Berthier:2017kkn,Duffell:2018iig,Lamb:2018ohw,Lamb:2018qfn,Mooley:2018qfh,Mooley:2018clx,Wu:2018bxg}, marked the beginning of multimessenger astronomy, i.e. the observation of coincident GWs with electromagnetic radiation.
This single event provided: i) the most direct evidence that BNSs are progenitors of the central engine that power short gamma-ray bursts (sGRBs); ii) a strong observational support to theoretical proposals linking stellar compact binary mergers with production sites for r-process nucleosynthesis and kilonovae~\citep{Li:1998bw,Metzger:2016pju, Troja:2017,Kasen:2017}; iii) an independent measure for the expansion of the Universe~\citep{lvk_hubble:2017,Dietrich:2020efo}; and iv) tight constraints on the equation of state (EoS) of matter at supranuclear densities~\citep{Rezzolla:2017aly,Ruiz:2017due,Shibata:2017xdx,Margalit:2017dij,lvk_eos:2018,lvk_eos:2019}.

GWs from both the inspiral and the postmerger epochs carry signatures of the EoS. During the inspiral, 
tidal forces transfer energy and angular momentum from the orbit to the NS. Most of the transferred energy is converted into gravitational radiation and deformation of the NS structure, while internal heating remains minimal. As a result, NSs are often considered to have effectively zero temperature during the inspiral,~see~e.g.~\citep{Kastaun:2016yaf}. 
GW observations of this phase, along with semi-analytical approaches~\citep{Flanagan:2007ix,Baiotti:2010xh} have been used to infer EoS-independent, quasi-universal relations that can be used to estimate NS properties. In particular, it has been shown that frequencies at GW peak amplitude  are tightly correlated with the tidal deformability of the two stars~\citep{Read:2013zra,Takami:2014tva,Rezzolla:2016nxn,Maione:2017aux}.

In contrast, during the postmerger phase, full general relativistic simulations have shown that the remnant can reach temperatures as high as $\approx 100\,\rm MeV$~\citep{Kastaun:2016yaf,Sun:2022vri,Bamber:2024kfb}, and hence a proper treatment of finite-temperature effects is required to extract signatures of the EoS during this epoch. Several numerical studies have assessed the properties of the GW spectrum which can be used to constraint the EoS ~\citep{Takami:2014tva,Bauswein2012,Stergioulas:2011gd,Hotokezaka:2013iia,Bauswein:2012ya,Takami:2014zpa}, to probe the relevance of thermal effects on the stability of the remnant~\citep{Rezzolla,Alford2018,Most2022,Miravet-Tenes:2024vba,Villa-Ortega:2023cps,Rivieccio_2024} and to assess its detectability. Recent studies suggest that thermal effects my be measurable with third generation detectors if the cold EoS is well constrained~\citep{Raithel:2023gct}.

Such GW observation can thus provide an understanding of the thermal properties of dense matter. A first-principles understanding of hot nuclear matter
properties requires the solution of the nuclear many-body 
problem at finite temperature~\citep{Rios2020,Brady2021,Keller2021,Keller2023}. 
This solution is challenging for several reasons, including the
uncertainties associated to the strong nuclear force and 
the complexity of solving the finite-temperature many-body Schr\"odinger equation~\citep{Drischler2021}. 
The notoriously difficult nature of the problem has motivated the use of
simplifying assumptions to incorporate thermal effects in hydrodynamical simulations. 
A widely used approach is the so-called \emph{thermal index} (or Gamma-law EoS), which relates the thermal components of energy density and pressure in the dense matter of NS interiors~\citep{Bauswein2012,Rezzolla}. 

In this approach, one decomposes the \textit{cold} and the \textit{thermal} components of the pressure and the energy density of the fluid,
\begin{align}
\label{eq:press_descomposada}
    P(n_b, Y_p, T)=&P(n_b, Y_p, T=0)+P_{\rm th}(n_b, Y_p, T) \, ,\\
    \label{eq:ener_descomposada}
    \epsilon(n_b, Y_p, T)=&\epsilon(n_b, Y_p, T=0)+\epsilon_{\rm th}(_bn, Y_p, T) \, ,
\end{align}
where $n_b$ is the number of particles per unit of volume, $Y_p = n_{proton}/n_b$ is the proton fraction and $T$, the temperature of the system. 
Thermal effects in the EoS can then be recast in terms of a thermal index, which depends only on the thermal pressure and thermal energy density~\citep{Rezzolla,Constantinou2015,Carbone2019,Kochankovski2022},
\begin{align}
    \Gamma_{\rm th} = 1 + \frac{P_{\rm th}}{\epsilon_{\rm th}}.
    \label{eq:th_index_general}
\end{align}
This is a dimensionless quantity that provides a simple characterization of the thermal properties of dense matter. 
Knowledge of $\Gamma_{\rm th}$ immediately allows for the thermal components of the EoS to be linked to one another. This can facilitate
BNS simulations by avoiding costly extrapolations of the EoS across density, temperature and isospin asymmetry space. 

There are a few simple, non-interacting models that provide predictions for the thermal index of dense matter, and indicate that it should be a constant~\citep{Rezzolla}. When nuclear interactions and other effects, like clustering or hyperon production, are accounted for, the density, temperature and isospin asymmetry dependence of $\Gamma_{\rm th}$ is more difficult to characterize~\citep{Constantinou2015,Carbone2019,Raduta_2021,Kochankovski2022}. Here, we aim to clarify the origin of some of these dependences by exploiting the virial expansion~\citep{Huang,Horowitz06_1,Horowitz06_2,Horowitz06_3}. 
The virial expansion allows for the quantification of nuclear interaction effects in a model-independent way. It has been used in the past to treat relevant astrophysical environments, including also the propagation of neutrinos in supernovae~\citep{Neutrino_virial_horowitz, Neutrino_virial_horowitz2}. This very useful tool, however, has so far not been exploited in the analysis of the thermal index of NS matter. 

Our analysis focuses in the high-temperature, low-density regime where the virial expansion converges, but also where simulations indicate that a significant portion of matter occurs in BNS mergers~\citep{Villa-Ortega:2023cps,Kastaun:2016yaf,Shibata:2019wef}.

As shown by \cite{Neutrino-beta-eq}, neutrinos can significantly influence the composition of matter, helping maintain it near $\beta$-equilibrium with minimal deviation. 
The regime where the virial expansion converges is such that thermal contributions are dominant and interaction effects are relatively small. 

In our analysis, we shall largely ignore nuclear clusters. It is well known that density, temperature and isospin asymmetry strongly affect nuclear clusterization. At sufficiently large temperature ($T \approx 15$) and isospin asymmetries, nuclei tend to dissolve completely~\citep{HS_EoS,Shen2011,Shen2020}.  The average temperatures in the bulk of the binary remnant following merger are typically above this threshold~\citep{Villa-Ortega:2023cps,Kastaun:2016yaf,Shibata:2019wef}, and hence it is reasonable to discard clustering treatment as a first approximation. 

The reminder of the paper is organized as follows. Sec.~\ref{section:virial} provides a comprehensive review of the key aspects of the virial expansion and the thermal index in the context of pure neutron matter (PNM). In Sec.~\ref{section:asymmetric}, we extend this analysis to more realistic scenarios by considering isospin-asymmetric matter consisting of neutrons and protons, along with the contributions from electrons and photons ($npe\gamma$ matter). 
Our results are compared to those of existing equations of state (EoSs) from the CompOSE database~\citep{compose1,compose2} in~Sec.~\ref{sec:comparison}. Finally, we summarize our results and conclude in Sec.~\ref{sec:conclusion}.  For reference, we present the numerical values of virial coefficients at different temperatures in Appendix~\ref{sec:appendixA} and discuss details of the virial expansion in asymmetric matter in Appendix~\ref{sec:appendixB}.


\section{Virial expansion for pure neutron matter}
\label{section:virial}

We begin our discussion by providing a brief description of the virial expansion in PNM. This is an informative exercise that provides an insight on the relevance of nuclear interactions for the thermal index in neutron-star matter. PNM is also ideally suited for the quantification of the truncation error in the virial expansion.

%
\subsection{Virial expansion}
\label{Sec:virrial_exp}
The virial expansion is a model-independent framework that accounts for strong interactions between nucleons in the EoS of a hot, dilute gas~\citep{Horowitz06_1,Horowitz06_2,Huang}.
For a single species, here assumed to be neutrons, the virial EoS is derived by expanding the pressure
in a power series of the fugacity $z=e^{\mu/T}$ as~\citep{Huang} 
\begin{align}
    P=\frac{2T}{\lambda^3}\sum^{\infty}_{n=1}b^{(n)}z^n \, ,
    \label{eq:press_general}
\end{align}
where $\mu$ is the chemical potential and $T$, the temperature of the system. Throughout this work, we set
the Boltzmann constant~$k_B=1$.  We also introduce the neutron de Broglie thermal wavelength 
$\lambda=\hbar\,(2\pi/m_n\,T)^{1/2}$, with $m_n$ the neutron mass. The term $b^{(n)}$  denotes the dimensionless virial coefficient at the $n$th order~\citep{Huang}. This expansion is valid under the condition that $z \ll 1$, which implies a regime of very low densities and high temperatures~\citep{Horowitz06_1}.
Our study focuses on the EoS and the thermal index under conditions of low density and high temperature,  consistent with the requirements of the virial expansion. We assume a homogeneous gas, focusing on a temperature range of approximately $T \approx 1-50$ MeV for temperature and a density range of $n_b \approx 10^{-5}-10^{-1}$ \, \rm {fm}$^{-3}$. To ensure that the expansion remains perturbative, we impose the condition of $z<0.3$ throughout our study. 

In the virial expansion framework, the $n$th virial coefficient $b^{(n)}$ encodes information about interactions and correlations between clusters of $n$ particles~\citep{Huang,virial_in_cond_matter}. These coefficients quantify how the presence of interactions between multiple particles affects the thermodynamic properties of the system, such as pressure or energy density. The first virial coefficient, $b^{(1)}=1$, corresponds to the behavior of non-interacting particles, while higher-order coefficients account for progressively more complex many-body correlations~\citep{HouDrut20,Hou_2020}.
Virial coefficients are typically split into two components, $b^{(n)}=b_0^{(n)}+\Delta b^{(n)}$. The non-interacting term, $b_0^{(n)}=(-1)^{n+1}/n^{5/2}$, is associated to the free Fermi gas. In contrast, the interaction-induced term $\Delta b^{(n)}$ accounts for $n$-body correlations such as  clustering, that may become significant under the relevant thermodynamic conditions~\citep{HouDrut20,Hou_2020}.
In fermionic systems, only the first few terms of the expansion can and have been computed~\citep{HouDrut20,Hou_2020}. 
Notably, these coefficients depend only on temperature~\citep{Huang,Horowitz06_1}. However, the overall thermodynamic quantities also depend on the fugacity $z$, which introduces an additional dependence on the chemical potential $\mu$ or, equivalently, the number density of the system $n$.

We start by considering PNM as a first approximation for NS matter, which enables us to analyze the effects of nucleon interactions. In nuclear physics, the virial expansion is typically truncated at second or third order, and the expressions for the EoS are well known~\citep{Horowitz06_1,Horowitz06_2,Horowitz06_3}. We reproduce them here for convenience, and to illustrate how they affect the thermal index. We work up to third order in the fugacity $z$, so the pressure can be expressed as
\begin{align}
P = \frac{2T}{\lambda^3}\left(z + z^2 b^{(2)}_n + z^3 b^{(3)}_n\right).
\label{eq:press_neutrons}
\end{align}
$b^{(2)}_n$ and $b^{(3)}_n$ are the second and third virial coefficients for PNM, respectively.
The density is obtained by differentiating the pressure with respect to the chemical potential which results in
\begin{align}
n_b =\frac{z}{T}\,\left(\frac{\partial P}{\partial z} \right)_{V, T}
=
\frac{2}{\lambda^3}\left(z + 2z^2 b^{(2)}_n + 3z^3 b^{(3)}_n\right). 
\label{eq:dens_neutrons}
\end{align}
The entropy density can be obtained by differentiating the pressure with respect to temperature
\begin{align}
s = \left(\frac{\partial P}{\partial T} \right)_{\mu}=
\frac{5}{2}\,\frac{P}{T} -n_b\,{\rm log} z +\frac{2}{\lambda^3}\left(z^2\, \overline{b_{n}^{(2)}} + z^3\, \overline{ b_{n}^{(3)} }\right),
\end{align}
where we introduce a bar notation,
$\overline{b_{n}^{(m)}} = T b_{n}^{(m)\prime}  = T \partial b^{(m)}_{n}/\partial T $, to denote the dimensionless temperature derivative of the $m^\text{th}$-order virial coefficient. 
Finally, the energy density is calculated from the entropy density and the pressure, $\epsilon= Ts+n_b\mu-P$, leading to
\begin{align}
    \epsilon=\frac{3}{2}P +\frac{2T}{\lambda^3}\left(z^2 \overline{ b_n^{(2)} }+z^3 \overline{ b_n^{(3)} }\right).
    \label{eq:ener_neutrons}
\end{align}

These expressions help clarify how nuclear interactions,
parametrized by the virial coefficients, 
change the thermal index of PNM. First, we note that
the virial expansion predictions for the pressure and the
energy density are intrinsically
thermal. In other words, the expressions vanish at $T =
0$,  e.g.~$P (n, T = 0) = 0$
and, by virtue of Eq.~(\ref{eq:press_descomposada}),
the pressure in Eq.~(\ref{eq:press_neutrons}) is the full thermal contribution, $P = P_{\rm th}$. The same holds for the energy density. 

Second, the pressure and energy density expressions in Eqs.~(\ref{eq:press_neutrons}) and~(\ref{eq:ener_neutrons}) 
can help us identify deviations from the behavior of a standard ideal gas, particularly departures from the conventional relation
$\epsilon=3/2P$. For PNM, this ideal gas law leads to a temperature and density independent thermal index, $\Gamma_{\rm th}=5/3$. We note in passing that 
a gas of free relativistic particles would have a thermal index $\Gamma_{\rm th}=4/3$.
Taking the ratio of the pressure and energy density in the virial approximation, we find an expression for the thermal index of PNM that reads
\begin{align}
\Gamma_\text{th} = \frac{5}{3} 
- \frac{4}{9} \frac{ 
\overline{b_n^{(2)}} + z \overline{b_n^{(3)}}
}{
1+ z \left( b_n^{(2)} + \frac{2}{3} \overline{b_n^{(2)}} \right) 
+ z^2 \left( b_n^{(3)} + \frac{2}{3} \overline{b_n^{(3)}} \right) 
}.
\label{eq:th_PNM_fullz}
\end{align}
This equation suggests that deviations from
the simple $5/3$ law will only occur when the derivative coefficients
$\overline{ b_{n}^{(2)}}$ and 
$\overline{ b_{n}^{(3)}}$ are large enough. The actual virial coefficients $b_{n}^{(m)}$ appear on the denominator with higher powers of fugacity and hence provide 
higher-order  corrections. As we shall see later explicitly, the temperature derivatives coefficients are in fact the main source of the deviations from the trivial non-interacting behavior of the thermal index in PNM.

There is, in fact, a known scenario where the thermal index is also a trivial constant but the system is strongly interacting. 
This scenario occurs for the unitary Fermi gas: a system of spin$-1/2$  fermions interacting only through an $S-$wave potential with infinite scattering length~\citep{HoMueller04,Horowitz06_1,LeeSchaefer06,HouDrut20}. For this system, the second order virial coefficient $b^{(2)} \neq 0$ is non-zero, but it is a temperature-independent constant. In this case, $b^{(2)\prime}=0$, and the thermal index reduces to the trivial value $\Gamma_{\rm th}=5/3$. PNM can be relatively well approximated as a unitary gas and, as we shall see below, the temperature dependence of the virial coefficients is very mild. One may thus anticipate that the thermal index of PNM may not be substantially affected by interactions.

\subsection{Virial coefficients}
We now provide a discussion on the calculation of the virial coefficients of PNM, which follows closely the scheme devised by~\cite{Horowitz06_1}. 
The second virial coefficient $b^{(2)}_n$  is entirely determined by the two-body neutron-neutron ($nn$) elastic scattering phase shifts~\citep{Huang}. The interacting term $\Delta b^{(2)}_n$ is calculated as
\begin{align}
\Delta b^{(2)}_n(T)=\frac{1}{\sqrt{2}\pi T} \int^{\infty}_0 \! dE \, e^{-E/2T} \delta^{tot}(E).
\label{eq:b2_n}
\end{align}
For PNM, $\delta^{tot}(E)$ is the sum of the isospin-triplet ($I=1$) elastic scattering phase shifts at a given laboratory energy, $E$,
\begin{align}
    \delta^{tot}(E)&=\sum_{S,L,J}(2J+1)\delta_{^{2S+1}L_J}(E) \\
    &= \delta_{^{1}S_0}+\delta_{^{3}P_0}+3\delta_{^{3}P_1}+5\delta_{^{3}P_2}+5\delta_{^{1}D_2}+\cdots\,. \nonumber
\end{align}
The sum runs over all partial waves allowed by spin statistics, with two-particle spin $S$ and angular momentum $L$, and includes a degeneracy factor  that depends on  the total angular momentum $J$. 

To compute $ \delta^{tot}(E)$, we use the phase-shifts from the Granada database with $J \leq 7$ for energies up to $350$ MeV~\citep{granada}. For energies above this threshold, we approximate the phase shifts as a constant. This approximation introduces an error of less than $3\%$ in the virial coefficients at the highest temperatures we consider ($T=50$ MeV).
We estimate the integral of Eq.~(\ref{eq:b2_n}) using numerical quadratures, e.g.~the QUADPACK package in scipy \citep{2020SciPy-NMeth}. 
The dimensionless virial coefficient derivatives $\overline{ b_{n}^{(m)}}$ are explicitly computed directly from their integral forms, rather than relying on numerical temperature derivatives. 

Since $nn$ scattering phase-shifts are not experimentally available, we estimate the $nn$ phase shifts by applying charge-independence breaking (CIB) corrections, as originally proposed by~\cite{Horowitz06_1}.     
We employ the effective range expansion
\begin{align}
    \frac{p}{\tan{\delta^{\tau \tau'}_{^{1}S_0}}}=-\frac{1}{a^{\tau \tau'}} + \frac{1}{2} r^{\tau \tau'} p^{2},
    \label{eq:eff_range_exp}
\end{align}
which is valid for the scattering of pairs of neutrons, $\tau=n$, or protons, $\tau=p$. The expansion
parametrizes the scattering amplitude information at low scattering momenta $p$ in terms of 
two length scales, the so-called scattering length $a^{\tau \tau'}$ and effective range $r^{\tau \tau'}$~\citep{Taylor_scattering}. 
For neutron-proton ($np$) scattering, we employ the values 
$a^{np} = -23.71\,\rm{fm}$ and 
$r^{np} = 2.75 \,\rm{fm}$ from~\cite{Hackenburg06}
in Eq.~(\ref{eq:eff_range_exp})
to compute an approximate phase-shift $\delta^{np}_{^{1}S_0}$ in the $^{1}S_0$ partial wave. 
We subtract this approximate expression to the Granada $np$ phase-shift 
and add up an approximate $nn$ phase-shift based on Eq.~(\ref{eq:eff_range_exp}) with a $nn$ scattering length value of
$a^{nn} = -18.5\,\rm{fm}$~\citep{Gardestig09}
and an effective range, $r^{nn} = 2.86 \,\rm{fm}$~\citep{Malone22}. 
This procedure yields results consistent with~\cite{Horowitz06_1}, with deviations within $1\%$ up to $200\,\rm MeV$ and $4\%$ up to $350\,\rm MeV$ in the total phase shift. 
We note that our analysis includes partial waves up to $J=7$, while~\cite{Horowitz06_1} used waves up to $L \leq 6$. 
For reference, we provide a table of virial coefficients and their derivatives in Appendix~\ref{sec:appendixA}. The code used to 
generate this data is available in~\citep{Rivieccio_Virial-EoS_2025}.

\begin{figure}
\includegraphics[width = \linewidth]{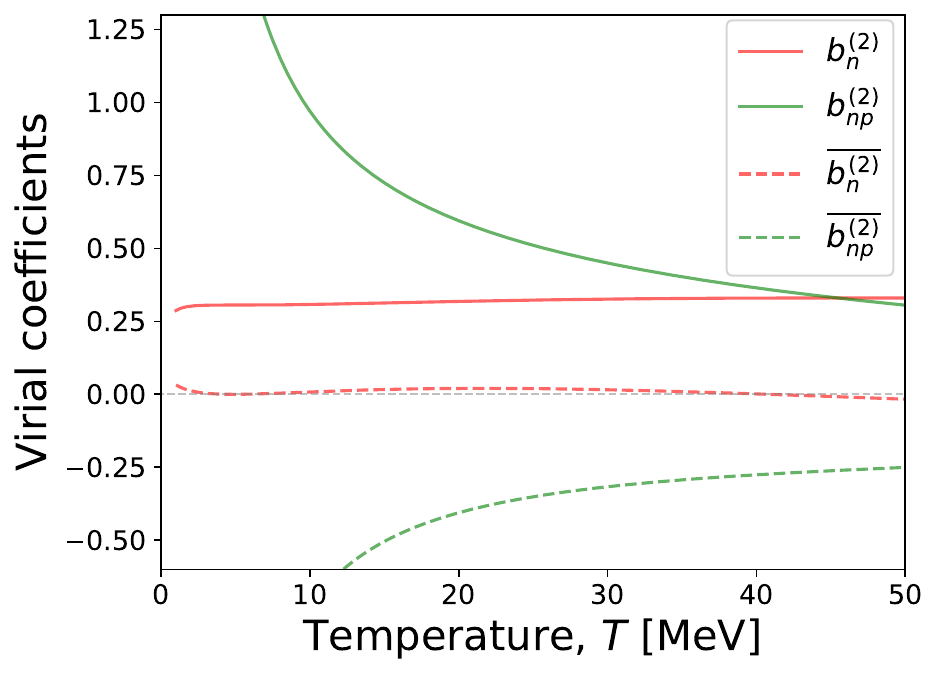}
\caption{\label{fig:Virial_Coeffs}
The second-order virial coefficient as a function of temperature for neutrons is shown by the solid red line, computed from Eq.~(\ref{eq:b2_n}). For comparison, the corresponding $np$ coefficient from Eq.~(\ref{eq:b2_np}) is displayed as a solid green line. The dimensionless temperature derivatives of these coefficients are represented by dashed lines in the same respective colors. The gray horizontal dashed line highlights the value of zero.
}
\end{figure}

Fig.~\ref{fig:Virial_Coeffs} displays the second-order virial coefficient $b^{(2)}_n$ as a function of temperature for PNM with a red solid line. 
Clearly, this virial coefficient is rather insensitive to temperature. As discussed by~\cite{Horowitz06_1}, this bodes well with the idea that PNM closely resembles a unitary Fermi gas, for which $b^{(2)}_n \approx 0.52$~\citep{HoMueller04,LeeSchaefer06,Horowitz06_1}.  
For PNM, in contrast, the value is closer to  $b^{(2)}_n \approx 0.3$, but $b^{(2)}_n$ remains essentially temperature independent. 

The level of temperature independence of the virial coefficient can be further quantified by examining the dimensionless temperature derivative coefficient ~$\overline{  b_n^{(2)}}$, which are displayed as red dashed line  in  Fig.~\ref{fig:Virial_Coeffs}. For PNM, these 
derivative remain quite small, $\overline{  b_n^{(2)}}<0.04$, across the relevant domains of temperature. As explained above, this suggests that the thermal index of PNM in the virial approximation should be very close to the non-interacting value. 
We also note that 
$\overline{  b_n^{(2)}}$ changes from a positive to negative sign around $T \approx 41$ MeV, a feature that has some significance on $\Gamma_{\rm th}$ as we shall see below.

Virial coefficients beyond second order are notoriously challenging to compute.  While some coefficients have been calculated for specific condensed matter systems, such as the unitary Fermi gas~\citep{Bedaque03,Liu09,Leyronas11,Kaplan11,Hou_2020,HouDrut20}, there are, to the best of our knowledge, no available results for $b^{(3)}_n$ in nuclear physics.
Following~\cite{Horowitz06_1,Horowitz06_2}, we instead employ the third-order virial coefficient to provide an error estimate for our results. 
We study variations of the third-order terms in order to  quantify the uncertainties associated to the 
truncation in the virial expansion. 
We assume that the ratio between the interacting components of the second- and third-order virial coefficients in PNM is analogous to that of the unitary Fermi gas~$\Delta b^{(3)}_{unit}/\Delta b^{(2)}_{unit}=-0.5022$~\citep{HouDrut20}.
Using this assumption, we estimate the third-order coefficient as
\begin{align}
    b^{(3)}_n=3^{-5/2}+\Delta b^{(3)}_n \approx 3^{-5/2}-0.5022\,\Delta b^{(2)}_n\,,
    \label{eq:b3n}
\end{align}
and the corresponding derivative coefficient as 
$\overline{ b^{(3)}_n } \approx -0.5022\,\overline{ b^{(2)}_n }$.
In the following, we shall present  second-order results as central values, and provide  uncertainty bands that reflect the inclusion of third-order virial coefficients, ranging from $+b^{(3)}_n$
and $+\overline{ b^{(3)}_n} $ to $-b^{(3)}_n$ and $-\overline{ b^{(3)}_n}$ in 
Eq.~\eqref{eq:th_PNM_fullz}.

\subsection{Thermal index}
Having determined the EoS from the virial expansion and having computed a set of virial coefficients, we can now explore the predicitions of the  virial expansion for the thermal index, $\Gamma_{\rm th}$. As explained above, 
it is evident that the virial expansion, by its very nature, predicts only the thermal components of the pressure and energy density.
Taking the pressure in Eq.~(\ref{eq:press_neutrons}) and the energy density in Eq.~(\ref{eq:ener_neutrons}) and employing Eq.~(\ref{eq:th_index_general}), we find an expression for the virial expansion of $\Gamma_{\rm th}$ itself. 
The expansion up to second order can be written as
\begin{align}
    \Gamma_{\rm th} = \frac{5}{3} -\frac{4}{9}  \Gamma^{(1)}z + \frac{4}{9}  \Gamma^{(2)}z^2,
    \label{eq:th_index_neutrons}
\end{align}
with first- and second-order coefficients given by
\begin{align}
\label{eq:gammas}
\Gamma^{(1)} &= 
\overline{  b_n^{(2)}} , \\
\Gamma^{(2)} & = \overline{  b_n^{(2)}}
\left[b_n^{(2)} + \frac{2}{3} \overline{  b_n^{(2)}}^2 \right] .
\end{align}
These expressions are interesting for a variety of reasons. First of all, in the non-interacting case, 
$\overline{ b_n^{(m)}}=0$, 
the thermal index is $\Gamma_{\rm th}^{(0)} = {5}/{3}$, as expected for a very dilute gas of non-relativistic neutrons. 
Second, the coefficients $\Gamma_{\rm th}^{(m)}$ in the virial expansion with $m=1$ and $2$ are globally multiplied by the $m=2$ dimensionless temperature derivative, $\overline{ b_{n}^{(2)} }$. For a strongly-interacting unitary Fermi gas, for instance, in which $b_n^{(m)} \neq 0$ but $\overline{ b_n^{(2)}}=0$~\citep{Hou_2020,HouDrut20}, the thermal index is exactly the same as for the non-interacting case. 

Furthermore, Eq.~(\ref{eq:th_index_neutrons}) indicates that $\Gamma_{\rm th}$ should approach the non-interacting limit for a very dilute 
gas of non-relativistic neutrons, $\Gamma_{\rm th}\rightarrow {5}/{3}$ as $z\rightarrow0$. Moreover, as we have already 
mentioned, $\overline{ b_{n}^{(2)} }$ is a very small and positive quantity up until about $T \approx 41$ MeV. In consequence, 
the first order correction $\Gamma^{(1)}$ is expected to be small and to reduce the thermal index with respect to the non-interacting baseline so long as $T < 41$ MeV. We give the values of $\Gamma^{(1)}$ in Appendix~\ref{sec:appendixA} for further reference.

\begin{figure}
 \includegraphics[width=\linewidth]{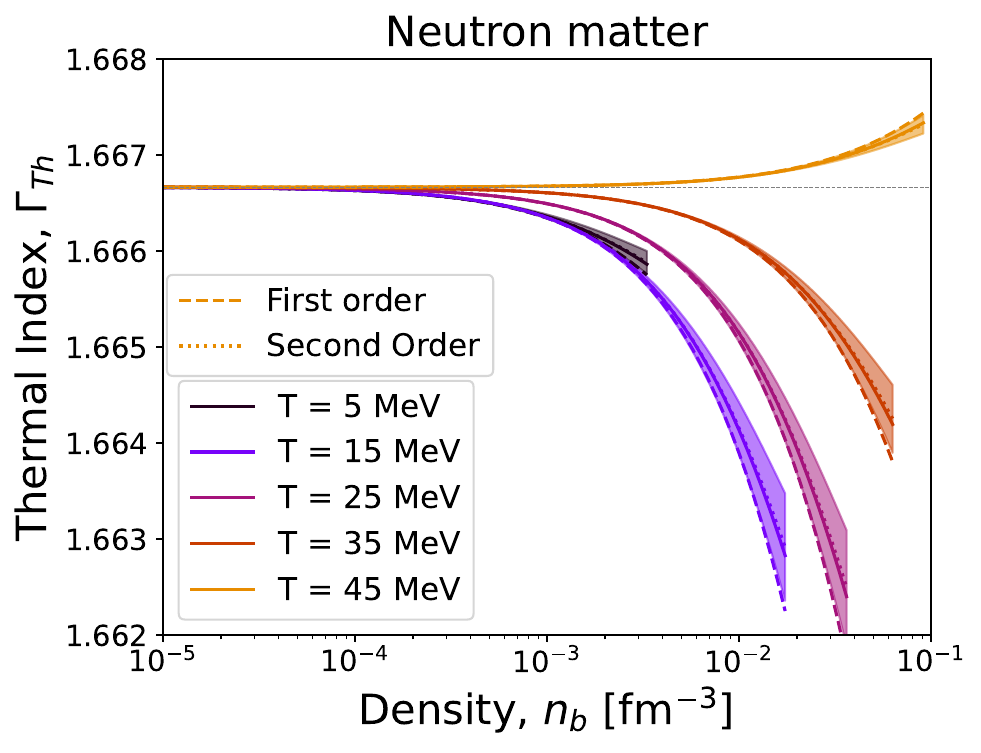}
 \caption{\label{fig:Gamma_neutron} Thermal index of neutron matter as a function of density for different temperatures. We show results from $T=5$ to $T = 45$ MeV with increments of $10$ MeV. Dotted (dashed) lines represent the first- (second-)order expansion of Eq.~(\ref{eq:th_index_neutrons}). Solid lines are the results of the ratio of Eq.~(\ref{eq:th_index_general}) with
 second-order pressure in Eq.~(\ref{eq:press_neutrons}) and energy densities in Eq.~(\ref{eq:ener_neutrons}).
 The shaded area shows an estimate for the error in the expansion, based on incorporating the third-order correction 
 $\pm b^{(3)}_n$. The dotted grey shows the value of $\Gamma_{\rm th} = {5}/{3}$ in the non-interacting limit, for reference. We only show results with fugacities below $z<0.3$.}
\end{figure}

Fig.~\ref{fig:Gamma_neutron} displays the thermal index $\Gamma_\text{th}$ of PNM as a function of density for temperatures ranging from  $T = 5$ MeV to $T = 45$ MeV, in increments of $10$ MeV. 
Each line illustrates the evolution of the thermal index for a given temperature. Dashed lines represent the first-order results using only $\Gamma^{(1)}$ and putting the quadratic contribution to zero, while the second-order expansion, that uses also $\Gamma^{(2)}$, is indicated by dotted lines. These two expressions are derived from Eq.~(\ref{eq:th_index_neutrons}) for a given fugacity $z$, with the corresponding number and energy densities, Eqs.~(\ref{eq:dens_neutrons}) and (\ref{eq:ener_neutrons}), computed at the same order. These expansions are compared to the full expression (solid lines), Eq.~\eqref{eq:th_PNM_fullz}, calculated using the ratio 
of the virial thermal pressure and energy density,  Eqs.~(\ref{eq:press_neutrons}) and~(\ref{eq:ener_neutrons}), computed up to second order. 
We provide an error-band associated to the truncation error of the third order in the expansion, employing the values
$\pm b^{(3)}_n$ and 
$\pm \overline{ b^{(3)}_n }$
described above. We note that these error bands are almost as large as the difference between the 
first and second orders, and hence are likely to be an overestimation of the associated truncation error. Based on the size of these bands, we evaluate the truncation error to be $\approx 0.05\%$ as an average across different temperatures.

A key takeaway from Fig.~\ref{fig:Gamma_neutron} is that interactions have a negligible impact on the thermal index of PNM. 
The maximum deviation that we observe is always $\lesssim0.2\%$ away from the ideal gas limit, $\Gamma_{\rm th}=5/3\approx 1.6666$.  
We note that this result holds across several orders of magnitude in density. We stop our virial expansion simulations whenever $z = 0.3$, which for the highest temperature considered here corresponds to a density of $n_b \approx 0.1$ fm$^{-3}$. 
The virial expansions thus suggests that, within the low-density, high-temperature regime where the expansion is valid, the strong interaction does not play a significant role 
in dictating thermal effects for PNM. The extremely naive approximation $\Gamma_{\rm th}=5/3$ should work well within this regime. We stress that these predictions are model-independent and dictated purely by two-nucleon
phase-shifts. Additional effects associated to many-body correlations at higher densities can and, in general, do modify this behavior~\citep{Constantinou2015,Carbone2019,Raduta_2021,Kochankovski2022}. 

Moreover, we also find that interaction effects generally tend to decrease the thermal index with respect to the non-interacting value at temperatures below $T<40$ MeV. In this regime, the density dependence of $\Gamma_\text{th}$ 
is  monotonous and the thermal index decreases with density. 
In contrast, for higher temperatures, $\Gamma_\text{th}$ is larger than $5/3$, and increases with density. 
These results can be traced back to the temperature dependence of the $\Gamma^{(n)}$ coefficients. 
Whereas at low temperatures $\Gamma^{(1)}$ and $\Gamma^{(2)}$ are small and positive, we find that around $T \approx 41$ MeV these two coefficients change sign. As observed in Fig.~\ref{fig:Gamma_neutron}, 
the main driver of the density dependence  is the first-order term 
 $\Gamma^{(1)}$. In the temperature regime where $\Gamma^{(1)}>0$, the thermal index is lower than $5/3$ and the
 second order correction tends to bring the results closer to this value. When $\Gamma^{(1)}<0$, the 
 predictions for $\Gamma_\text{th}$ are above $5/3$ but the second-order effect tends to reduce further the thermal 
 index. Moreover, the onset of interaction effects, as measured by
a fixed deviation with respect to the non-interacting case, happens at higher densities as temperature increases. For 
instance, the value $\Gamma_\text{th}=1.665$ is reached at around $n_b \approx 6 \times 10^{-3}$ \, \text{fm}$^{-3}$ at $T=15$ MeV, but at $T=35$ MeV a density of $n_b \approx 4 \times 10^{-2}$ \, \text{fm}$^{-3}$ is required instead. 
We offer a more detailed analysis of the temperature dependence of the thermal index virial coefficients in Appendix \ref{sec:appendixA}.

Overall, our analysis indicates that the thermal index of PNM within the virial approximation is highly insensitive to temperature, density and interaction effects. We note that within the region of validity of the expansion, the results of the virial expansion are model-independent, and hence predictions for the PNM thermal index from any other theoretical methods should coincide with our findings. 

\section{Virial expansion for isospin asymmetric matter}
\label{section:asymmetric}

The results of PNM only provide a crude approximation for the matter that is encountered in BNSs. We extend our 
approach towards more realistic settings by considering isospin asymmetric matter formed of neutrons and protons,
together with the contribution of electrons and photons. Multicomponent $npe\gamma$ matter is expected to be a relevant for a 
substantial section of the phase space of BNS remnants, particularly at high temperatures in which nuclei are likely
to be photo-disintegrated~\citep{Neutrino-beta-eq}. 
We begin by introducing the nucleonic contributions within the virial approximation, and then move on to discuss  how the presence of leptons and photons changes the thermal index. 

\subsection{Virial expansion}

To obtain the virial EoS for arbitrarily asymmetric matter, we extend the procedure of Sec.~\ref{section:virial} to a  multicomponent case, including neutrons and protons 
following ~\citep{Horowitz06_2}. We expect the truncation error to be as small as in the PNM case. 
For conciseness, we neglect a full assessment of this error in $npe\gamma$ matter. In other words, we work at second order in the virial expansion only in this case. We provide mathematical and code implementation details in Appendix~\ref{sec:appendixB}. 

We work in conditions of low density and high temperature and assume a homogeneous gas. 
We impose the condition of small fugacities for neutrons and protons, $z_\tau<0.3$. 
In the regime of interest (densities around $n_b \approx 10^{-5}-10^{-1}$ \, \rm {fm}$^{-3}$ and temperatures $T \approx 1-50$ MeV), the only non-trivial cluster that we consider is the deuteron. This appears explicitly as an interacting term in the virial expansion, see Eq.~(\ref{eq:b2_np}). 

Compared to PNM, the virial expansion in the isospin asymmetric case depends on an additional coefficient, $b_{np}^{(2)}$, quantifying neutron-proton interactions. We display this virial coefficient in Table~\ref{TableAppenA}. We find that it reaches a maximum at low temperatures, driven by the deuteron bound state contribution. 
However, as the temperature increases, this contribution becomes negligible, which is physically expected because of the thermal disruption of the bound state. This trend is clearly illustrated in Fig.~\ref{fig:Virial_Coeffs}, where the virial coefficients for the scattering of $nn$ and $np$ are displayed by red and green solid lines, respectively, while their derivatives are shown as dashed lines. 
Whereas the $nn$ coefficients are rather temperature independent, the $\overline{ b^{(2)}_{np}}$ derivative is negative and decreases in absolute magnitude with temperature. 

Our analysis indicates that nuclear interaction effects, driven by the virial approximation, do not modify substantially the thermal index in the region of interest. We provide an illustration of this behavior in Subsec.~\ref{subsec:results}.

At low temperatures and levels of asymmetry, the formation of $\alpha$ particles and other clusters may affect the thermodynamical properties~\citep{Horowitz06_2,HS_EoS,Lattimer}. 
Clusterization is sensitive to the thermodynamical properties of matter.
Numerical simulations have shown that the proton fraction $Y_p$ in BNS merger remnants typically ranges between $0.1$ and $0.4$~(see~e.g.~\cite{Fujibayashi:2020dvr}), depending on the specific region. In the dense, hot core, where temperatures exceed 
$10\,\rm MeV$, the proton fraction remains low $Y_p\approx 0.1$ due to the neutron-rich environment. However, in the outer layers and the accretion disk, where densities are lower and weak interactions are more significant, $Y_p$ can increase to around $0.2-0.3$ (see e.g.~\cite{Radice:2018xqa,Sun:2022vri,Perego:2014fma}). In the phase space that we are interested in for BNS simulations, the $\alpha-$particle fraction is expected to be relatively small~\citep{Horowitz06_2,Lattimer,Shen2011}.
We therefore ignore the effects of clusters with $A>2$ in the following. 
While we observe that the presence of clusters and heavy nuclei is significant at relatively low temperatures, as we describe in Sec.~\ref{sec:comparison} in the context of a comparison with more sophisticated models, we argue in the following that the bulk dependence of the thermal index in temperature, density and proton fraction across the relevant phase space can already be understood in this simple homogeneous approximation.

\subsection{Lepton and photon contributions}
We treat electrons and photons as ideal gases. The EoS for the non-interacting gas of electrons follows that of an ideal Fermi gas. However, for temperatures $T>2 m_e c^2 \approx 1\,\rm MeV$, the contributions from electron-positron pairs must also be considered.
We assume equilibrium in the $\gamma\longleftrightarrow e^{+} + e^{-}$ reaction, leading to
\begin{align}
    \mu_{e^+}=&-\mu_{e^-} \, .
\end{align}
 For electrons, we define the degeneracy parameter $\eta ={\mu_e}/T$ and the relativity parameter $\theta={T}/{m_e\,c^2}$. For positrons, in contrast, the degeneracy parameter is instead $\kappa = -\eta -{2}/{\theta}$, due to the mass excess related to the pair production. With this, we obtain the expressions for the pressure and energy density of leptons~\citep{Gerald},
\begin{align}
P^{lep}=\frac{2\,\sqrt{2}\,m_e^4\,c^5\,\theta^{5/2}}{3\,\pi^2\,\hbar^3}
& \big[ F_{3/2}(\eta,\theta)+F_{3/2}(\kappa,\theta) \\
    &+\frac{\theta}{2}\left(F_{5/2}(\eta,\theta) 
    + F_{5/2}(\kappa,\theta)\right)\big], \nonumber \\
\epsilon^{lep}= \frac{\sqrt{2}\,m_e^4\,c^5\theta^{5/2}}{\pi^2\,\hbar^3} &
    \big[ F_{3/2}(\eta,\theta) + F_{3/2}(\kappa,\theta) \\ &+ \theta\,\left(F_{5/2}(\eta,\theta)+F_{5/2}(\kappa,\theta)\right)\big] \nonumber\\
    &+ \,m_e\,c^2\,n_{e^+} + m_e\,c^2\,n_{e^-}. \nonumber
\end{align}
We note the the last two terms account for the rest mass contribution of electrons and positrons. $F_k$ represents the generalized Fermi-Dirac integral, 
 \begin{align}
    F_k(\eta,\theta)=\int\limits_0^{\infty}
     dx\,\frac{x^k\sqrt{1+{\theta\,x}/{2}}}{1+e^{x-\eta}}.
\end{align}
We evaluate this integral numerically using the implementation of the QUADPACK package in the Scipy  library for Python \citep{2020SciPy-NMeth,Rivieccio_Virial-EoS_2025}. We then obtain the thermal pressure and energy density using the relations $P^{\rm lep}_{\rm th}=P^{\rm lep}-P^{\rm lep}(T=0)$ and $\epsilon^{\rm lep}_{\rm th}=\epsilon^{\rm lep}-\epsilon^{\rm lep}(T=0)$, where we employ the zero-temperature expressions,
\begin{align}
P^{\rm lep}(T=0)=\frac{m_e^4\,c^5}{24\,\pi^2\,\hbar^3} \big[&x_F\,\sqrt{1+x_F^2}\,\left(2\,x_F^2-3\right) \nonumber\\
&+ 3\, \arcsinh{x_F} \big] \, ,
&  \\
\epsilon^{\rm lep}(T=0)= \frac{m_e^4\,c^5}{8\,\pi^2\,\hbar^3} \big[&x_F\,\sqrt{1+x_F^2}\,\left(1+2\,x_F\right) \nonumber\\
&- \, \arcsinh{x_F} \big] , &
\end{align}
with the dimensionless Fermi momentum,  $x_F= \hbar\,\left(3\,\pi\,n_b\,Y_p\right)^{\nicefrac{1}{3}}/m_e c$.
We employ this formula only for the electrons, since at $T=0$ the pair production from photons is not possible, leading to an absence of positrons.

For the photon gas,  the pressure and energy density are given by~\citep{Cox2003,Lattimer},
\begin{align}
    &P^{\gamma}=\frac{\pi^2\,T^4}{45\,\hbar^3\, c^3}\,,    &\epsilon^{\gamma}=3P^{\gamma}.
\end{align}
These quantities are inherently thermal, with no corresponding cold counterparts.
By incorporating the contributions from leptons and photons, we calculate the thermal index using the thermal pressures and energy densities of all components, which is given by
\begin{align}
   \Gamma_{\rm th}=1+\frac{P_{\rm th}^{\rm nuc}+
   P_{\rm th}^{\rm lep}+P^{\gamma}_{\rm th}}{\epsilon_{\rm th}^{\rm nuc}+\epsilon_{\rm th}^{\rm lep}+\epsilon_{\rm th}^{\gamma}}\,.
   \label{eq:th_index_asymmetric}
\end{align}
In principle, the thermal index depends on the total baryonic density, $n_b$; the temperature, $T$; and the matter composition through the proton fraction $Y_p={n_p}/{n_b}$. The latter dependence can be simplified in $\beta$-equilibrium. 

%
\subsection{$\beta-$equilibrium}
Cold, isolated NSs are expected to achieve $\beta-$equilibrium due to the long timescales involved in their evolution, allowing particle interactions to balance the chemical potentials of neutrons, protons, and electrons. In addition, theoretical indications suggest that matter within BNS remnants may also be $\beta-$equilibrated (see e.g.~\cite{Fujibayashi:2020dvr}). However, the dynamics of the merger process can lead to rapid changes in density and temperature, disrupting the equilibrium established in isolated stars. Furthermore, the timescales for dynamically evolving matter in the interior of BNS may be shorter than the expected times for $\beta-$equilibration~\citep{Beloborodov2003,Yuan2005}.
Our virial expansion approach is capable of addressing both $\beta-$equilibrated matter and matter with arbitrary asymmetry. 

However, for clarity, we briefly review the equations relevant to $\beta-$equilibrium. In this regime, $\beta$ decay and electron capture reactions are expected to achieve chemical equilibrium~\citep{Misner,Shapiro,Haensel},
\begin{align}
    n&\rightarrow p+e+\Bar{\nu}_e,\\
    p&+e\rightarrow n+\nu_e.
\end{align}
This leads to a relationship between the  
chemical potentials of neutrons, protons, and electrons, 
\begin{align}
    \mu_n + m_n c^2=\mu_p + m_p c^2 +\mu_{e^-}+m_{e^-} c^2\,.
    \label{eq:beta_eq}
\end{align}
To maintain charge neutrality, the densities of protons, electrons, and positrons must satisfy the  constraint
\begin{align}
    n_p + n_{e^+}=n_{e^-}.
\label{eq:charge_neutrality}
\end{align} 
For each value of baryon density $n_b$ and temperature $T$, the previous equations fix the proton fraction, $Y_b(n_b,T)$, at equilibrium. We show below a few indications of the $\beta-$equilibrium trajectory in the relevant domain of densities and temperatures. Broadly speaking, at very low densities, our approach predicts relatively proton-rich matter, with $Y_p \approx 0.5$, whereas matter becomes neutron-rich close to nuclear densities.

\subsection{Numerical results}
\label{subsec:results}

\begin{figure}
\includegraphics[width=\linewidth]{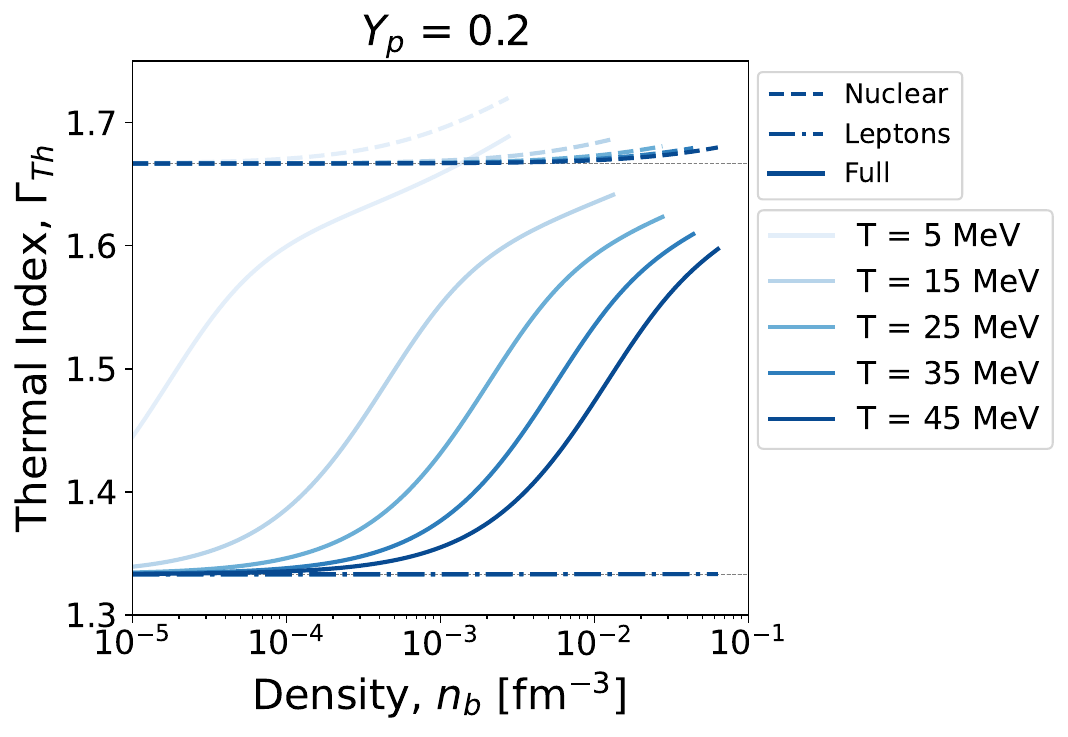}  
\caption{\label{fig:Gamma_contributions} 
Thermal index of $npe\gamma$ matter as a function of baryon density $n_b$ for different temperatures and a fixed proton fraction of $Y_p=0.2$. Solid lines show the total thermal index of asymmetric matter, Eq.~\eqref{eq:th_index_asymmetric}. Dashed lines display the nuclear thermal index, Eq.~\eqref{eq:th_nuc}, while the dashed-dotted line represents the lepton contribution. The two dotted lines indicate the thermal index of an ideal relativistic and non-relativistic gas, corresponding to $\Gamma_{\rm th} = {4}/{3}$ and ${5}/{3}$, respectively.
} 
\end{figure}

We start our analysis of $npe\gamma$ matter by looking into the separate contributions of nucleons and leptons to the thermal index. We define the nuclear thermal index,
\begin{align}
\Gamma_\text{th}^\text{nuc}=1 + \frac{P^\text{nuc}_\text{th}}{\varepsilon^\text{nuc}_\text{th}} \, ,
\label{eq:th_nuc}
\end{align}
as the ratio of the nuclear-only contributions to the thermal and energy density of matter. In isospin asymmetric matter, the virial expansion expressions for the thermal index are relatively  complex, see \eqref{eq:th_asymmetric}. We show results for this quantity as a function of density for different temperatures in Fig.~\ref{fig:Gamma_contributions}. The nuclear thermal index is shown in dashed lines for different temperatures for $npe\gamma$ matter with a proton fraction of $Y_p=0.2$. The corresponding lepton thermal index,  $\Gamma^\text{lep}_\textrm{th}=1+P_\textrm{th}^\text{lep}/\varepsilon_\textrm{th}^\text{lep}$, is shown in dash-dotted lines. 

The nuclear thermal index at low densities always tends to the non-interacting value $\Gamma_\text{nuc}^\text{th} = 5/3$. As the density increases, we observe a small changes due to nuclear interactions. At $T=5$ MeV, the nuclear thermal index deviates from the free value at densities above $n_b \approx 10^{-4}$ fm$^{-3}$, and reaches a maximum value $\approx 1.72$ before the virial expansion breaks down. This corresponds to a deviation of $\approx 3 \%$ from the free value. At higher temperatures, we find that the deviation from the non-interacting value is delayed in density and, in fact, it is generally smaller. 

First, we stress the fact that, just as in PNM, the modifications due to the interaction within the virial expansion are small in the region where the expansion is valid. 
Interaction effects modify the nuclear thermal index by a maximum of $\approx 10\%$, at $T=5$ MeV, a proton fraction of $Y_p=0.5$ and $n_b \approx 6\times10^{-3}$ fm$^{-3}$. 
However, unlike PNM, we find that in isospin-asymmetric matter the nuclear thermal index increases with respect to the non-interacting value. This interaction-induced effect is reduced as temperature increases and can be partially understood from  Eq.~\eqref{eq:th_index_asymmetric}. The derivative virial coefficient $\overline{b^{(2)}_\text{nuc}}$ is always negative and at least an order of magnitude larger than $\overline{b^{(2)}_n}$ (see Table \ref{TableAppenA}). Therefore, one expects it to dominate the numerator in Eq.~\eqref{eq:th_index_asymmetric} whenever $z_p$ becomes relatively large (e.g. for moderate proton fractions). This turns the interaction contribution to the nuclear thermal index negative. The steep decreasing temperature dependence of $\overline{b^{(2)}_\text{nuc}}$ also suggests that medium modifications should be mitigated with temperature, as observed in Fig.~\ref{fig:Gamma_contributions}. While we do not explicitly assess the virial expansion truncation error of the thermal index in this case, we expect it to be as small or comparable to the PNM case in relative terms inside the range of convergence of the virial expansion.

We now turn our attention to the total thermal index, Eq.~\eqref{eq:th_index_asymmetric}, shown in solid lines in Fig.~\ref{fig:Gamma_contributions}. Compared to the nuclear thermal index, the density dependence of the total thermal index is much more pronounced. In all cases, we find that the total thermal index is a monotonically increasing function of density. 

At low densities (say, below $n_b \approx 10^{-4}$ fm$^{-3}$ for the $T=25$ MeV case), leptons dominate the pressure budget of matter and, in consequence, the thermal index of $npe\gamma$ matter is that of a relativistic free Fermi gas. 
In this region, the total thermal index is basically the same as the lepton thermal index, $\Gamma_\textrm{th}=\Gamma^\text{lep}_\textrm{th} \approx 4/3$.
We stress the fact that this lepton index is independent of density and temperature across the whole regime explored in our work, in spite of the fact that we explore a wide range of temperatures for these two lepton species (electrons and positrons). In contrast, at high densities  neutron degeneracy sets in and dominate the pressure and thermal index. We find, in agreement with the previous Section, that 
$\Gamma_\text{th}^\text{nuc} \approx {5}/{3}$, with a small effect of interactions that is visible only at relatively high densities. 
At $T=5$ MeV, the total thermal index shows a quasi-linear behavior beyond $n_b \approx 10^{-3}$ fm$^{-3}$, when the total index gets right above the non-relativistic value. This is an imprint of nucleon-nucleon interactions incorporated in the virial approximation. 
Our findings indicate that the total thermal index of $npe\gamma$ matter evolves smoothly with density, from a lepton-dominated region to the nuclear-dominated regime. 
A temperature-dependent inflection point is clearly visible between these two regimes. We now explore whether this smooth behavior depends on temperature $T$ and proton fraction $Y_e$.
 
\begin{figure*}
\includegraphics[width=\linewidth]{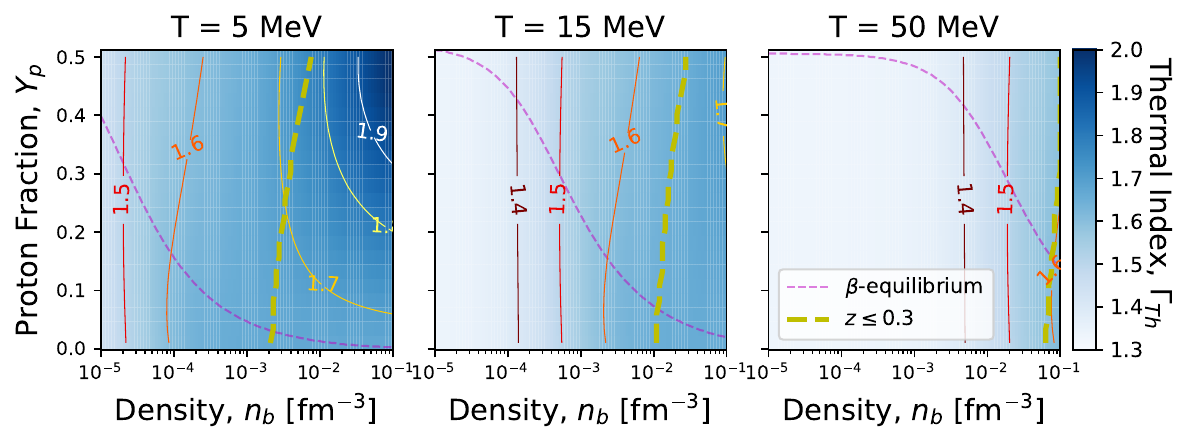}
\caption{\label{fig:panel} 
Density contour plot of the thermal index of $npe\gamma$ matter as a function of baryon density $n_b$ and proton fraction $Y_p$. The three panels correspond to temperatures $T=5$, $15$ and $50$ MeV from left to right. Contours are displayed in steps of $0.1$. The dashed violet line represents the $\beta-$equilibrium trajectory, while the yellow line indicates the constraint $z_n,z_p\leq 0.3$.  The virial approximation is valid to the left of this line.
}
\end{figure*}

In numerical relativity simulations, matter typically begins in a state of chemical equilibrium. Including neutrinos in the detailed balance is crucial, as their interactions help regulate weak processes such as $\beta$-decay which influence the evolving composition of matter. While the presence of neutrinos can help maintain the composition close to $\beta-$equilibrium, dynamical conditions during events like BNS mergers may temporarily disrupt this equilibrium~\citep{Neutrino-beta-eq}. 
Fig.~\ref{fig:panel} shows the full thermal index  $\Gamma_{\rm th}$  as a function of the proton fraction $Y_p$ ($y$-axis) and the baryon number density $n_b$ ($x$-axis). We choose three representative temperatures in the panels going from left to right: 
 $T=5$, $15$ and $50$ MeV. We choose contour lines between values of $\Gamma_{\rm th}=1.4$ and $1.9$, spaced in steps of $0.1$, to easily identify the evolution of the thermal index with density and temperature.
We also draw two dashed lines in each panel. 
The bold, yellow curve indicates the boundary of the region where $z_n,z_p\leq 0.3$. In other words, to the left of this curve the virial expansion should converge well and we expect our predictions to be robust. To the right of this curve, in-medium effects need to be considered beyond the virial approximation.

The three panels in Fig.~\ref{fig:panel} show remarkably similar behavior. The thermal index is roughly independent of the proton fraction in the region where the virial approximation is valid. At $T=5$, $15$ and $50$ MeV, the contour line $\Gamma_{\rm th}=1.5$ is almost vertical and is centered around $n_b \approx 2\times10^{-5}$ fm$^{-3}$; $5 \times 10^{-4}$ fm$^{-3}$; and $2\times10^{-2}$ fm$^{-3}$, respectively. The dependence on proton fraction is largest at lower temperatures, as indicated in the leftmost panel. 
There, the virial approximation is only valid up to $n_b \approx (2-8) \times 10^{-3}$ fm$^{-3}$. We find that, beyond those densities, there is a relatively complex dependence in $Y_p$. This prediction, however, cannot be taken as a robust prediction of our model since it falls beyond the convergence region of the virial expansion. 
In contrast, at higher temperatures (e.g. $T=50$ MeV on the right panel) the virial approximation is valid up to $0.1$ fm$^{-3}$, with the thermal index ranging between $\Gamma_{\rm th}\approx 1.3$ and $1.6$. Results in those conditions are essentially independent of proton fraction. 

The narrow, violet dashed curve in each panel displays  the trajectory in the $Y_p-n_b$ plane that corresponds to $\beta$-equilibrium. At low densities, we find that matter tends toward a symmetric composition ($Y_p \approx 0.5$) across all temperatures as expected from general physics arguments. At a fixed number density, increasing the temperature effectively raises $Y_p$. For instance, at $n_b=10^{-3}$ fm$^{-3}$, we find that the equilibrium proton fraction is $Y_p \approx 0.05$ at $T=5$ MeV; $0.3$ at $T=15$ MeV; and $0.48$ at $50$ MeV. A symmetric composition at low density has a substantial impact on the thermal index. Proton-rich matter is also lepton-rich due to charge neutrality, so that the thermodynamics of the low-density regime, which is the region where the virial approximation holds, is lepton dominated. In this situation, the thermal index remains close to the minimum $\Gamma_{\rm th}=4/3$. As density increases, the thermal index smoothly increases towards the nucleon-dominated, degenerate regime where $\Gamma_{\rm th} \approx 5/3$.

\subsection{Analytic parametrization}

The above observations suggest that it is possible to derive an analytical expression for the density evolution of the thermal index in $\beta-$equilibrium. The thermal index of $npe\gamma$ matter continuously grows from $\Gamma_{\rm th}=4/3$ to $5/3$ as a function of density. The transition between these two regimes is driven by the evolution of the $\beta-$equilibrium proton fraction, which in turn is relatively simple to model.
We use a heuristic formula to account for this transition, 
\begin{align}
\label{eq:approx}
    \Gamma_{\rm th}=\frac{4}{3}+\frac{1}{3}\frac{n_b}{n_b+n_{\rm inf}}\,,
\end{align}
with $n_{\rm inf}$ the density at which the regime shift mentioned above is expected to occur. 

We estimate the transition density considering that, at $n_{\rm inf}$, the condition $P^e_{\rm th}\approx P^{nuc}_{\rm th}$ is met. 
We approximate the thermal pressure of electrons as~\citep{Lattimer}
\begin{align}
P^e_{\rm th}= \frac{\mu_e^2\,T^2}{6\,\hbar^3\, c^3}\,,
\end{align}
and employ the relativistic, degenerate expression for the electron chemical potential, $\mu_e=(3 \pi^2 n_{e^-})^{1/3}$. Ignoring the presence of positrons, the electron density is $n_{e^-} \approx n_p= Y_p n_b$. For the nucleon thermal pressure, $P^{nuc}_{\rm th}$, we assume that at extremely low densities nucleons behave like a classical gas and hence $P^{nuc}_{\rm th} \approx n_b T$. 
Taking $Y_p\approx0.55$ as an illustrative value at very low densities, the density at the inflection point is 
\begin{align}
n_\textrm{inf} = 1.5 \times 10^{-4} 
\left( \frac{T}{10 \textrm{ MeV}} 
\right)^3 \,\textrm{fm}^{-3}.
\label{eq:infpoint}
\end{align}
This simple estimate suggests that: (a) the inflection point has a strong dependence on temperature; and (b) matter at saturation density may become thermally dominated by leptons at $T \approx 100$ MeV. 

\begin{figure}[t]
\includegraphics[width=\linewidth]{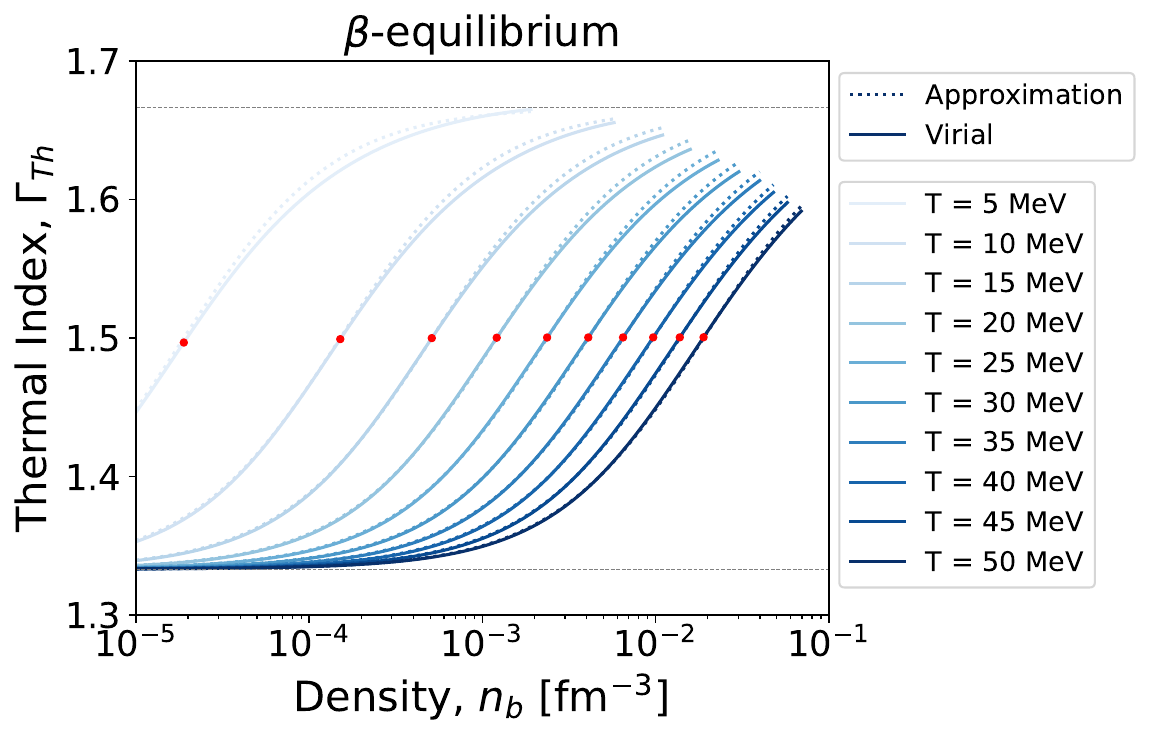}
\caption{\label{fig:Gamma_sigmoid}
Thermal index of $npe\gamma$ matter as a function of baryon density $n_b$ for temperatures ranging from $T=5$ to $50$  MeV in steps of $5$ MeV. For each temperature, a solid line represents the thermal index of asymmetric matter at  $\beta$-equilibrium, accounting for interactions through the virial approximation. The dotted line indicates the approximation of the thermal index as given in Eq.~(\ref{eq:approx}). Red dots mark the inflection densities of Eq.\eqref{eq:infpoint}. Gray dotted lines represent the values of $\Gamma_{\rm th}$ in both the relativistic and non-relativistic limits. All data shown are consistent with the condition $z<0.3$.
}
\end{figure}

Fig~\ref{fig:Gamma_sigmoid} displays a comparison between the heuristic approximation from Eq.~\eqref{eq:approx} and the full $npe\gamma$ thermal index at 
$\beta-$equilibrium. The density dependence of the heuristic formula is illustrated for temperatures ranging from  from $T=5$ to $50$ MeV (dotted lines) with the actual values of $\Gamma_{\rm th}$ (solid lines).
The red dot in each line indicates the position of the inflection density $n_{\rm inf}$ in Eq.~(\ref{eq:infpoint}) for which the heuristic formula gives $\Gamma_{\rm th}=1.5$, 
the value in between the lepton and the fermion-dominated regimes. 
Remarkably, the inflection density predicted by our simple model provides a very good approximation to the complete virial approximation numerical simulations. 
Within the selected range of densities and temperatures, our heuristic formula predicts the full thermal index with deviations of no more than 
$\lesssim 0.5\%$.    

We saw in the previous Section that the effect of interactions in the thermal index of neutron matter is negligible in the relatively wide range of densities and temperatures where the virial approximation holds. This is also the case for $npe\gamma$ matter. We do not show results here for brevity, but the effect of interactions in the temperature and density regime shown in Fig.~\ref{fig:Gamma_sigmoid} is very small. Although interactions can yield a $9\%$ ($10\%$) modification
in $P_\text{th}$ ($\varepsilon_\text{th}$), they only change $\Gamma_\text{th}$ by $0.3 \%$~\citep{Nadal_TFG}.

\section{Comparison with database-EoS} 
\label{sec:comparison}

Our findings indicate that the thermal index of $npe\gamma$ matter can
be understood employing relatively simple ideas in the low-density and high-temperature regime. Our analysis suggests that the thermal index is almost independent of proton fraction. 
Importantly, in the low density regime, the thermal index is dominated by leptons and hence also independent of density, with values very close to $4/3$.
Our approach does not account for the effects of clustering beyond the presence of deuterons and, therefore, may not fully capture every detail of the physical system.
We now turn to comparing our results to state-of-the-art EoSs as 
from a complete astrophysical database, CompOSE\footnote{https://compose.obspm.fr} ~\citep{compose2,compose1,compose3}.
These EoSs incorporate contributions from various phases of matter and, more importantly, from nuclear clusters. With this, we want to try and elucidate which effects, if any, we can expect from more realistic nuclear physics simulations. We shall examine this by looking at the nuclear-only thermal index, Eq.~\eqref{eq:th_nuc}. This is more sensitive to nuclear effects than the total thermal index, as we shall see in the following. 

\begin{figure}
    \centering
    \includegraphics[width=\linewidth]{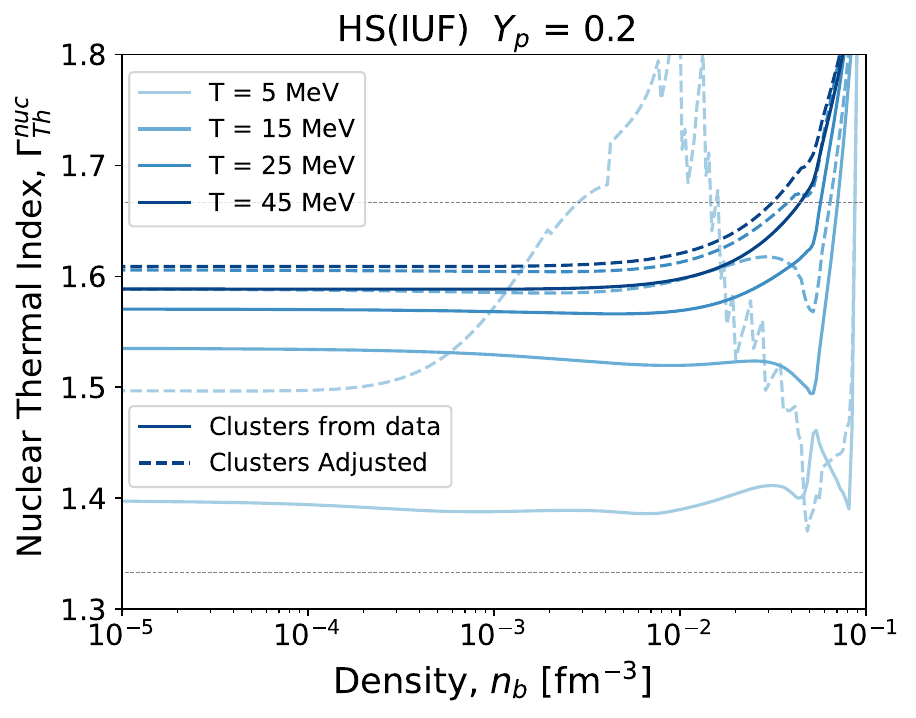}
    \caption{Nuclear thermal index Eq.~\eqref{eq:th_nuc} as a function of baryon number density for the HS(IUF) EoS taken from CompOSE database fixing $Y_p = 0.2$. The dashed line have been adjusted accounting for the binding energy while the solid not. The line color changes with temperature from $5$ MeV to $45$ MeV. }
    \label{fig:binding-energy}
\end{figure}

To derive the thermal index, we apply Eq.~(\ref{eq:th_index_general}) directly on the data provided in CompOSE. We employ the following EoSs because they include a variety of theoretical approaches of nuclear interactions and clustering effects: FYSS(TM1) \citep{FYSS_2011,FYSS_2013,FYSS2017}, HS(IUF) \citep{HS_EoS,IUF_EoS}, HS(TMA) \citep{HS_EoS, TMA}, LPB(chiral) \citep{LPB1,LPB2}, SFH(SFHo) \citep{SFH}, SFH(SFHx) \citep{SFH}, STOS(TM1) \citep{STOS1,STOS2} and TNTYST(KOST2) \citep{TNTYST}. In order to calculate the thermal index we need to isolate the thermal contributions subtracting the data slice at zero temperature to the one of interest, as explained previously.

The lowest temperature available for the tabulated EoSs employed is $0.1$ MeV. We use this value to compute the thermal EoS using Eqs.~(\ref{eq:press_descomposada}) and (\ref{eq:ener_descomposada}).
To verify the robustness of our results, we also calculated the thermal index using a minimum temperature  of  $T=0.2$  MeV.
The nuclear thermal index shows a slight offset with respect to the previous case in the low-density regime where the behavior is constant. At higher densities, it displays an increasing trend. This increase happens at a density that is roughly one order of magnitude smaller than the lowest temperature cases.
We deduce that the minimum temperature is critical in determining the thermal properties. Temperatures lower than $T=0.1 \, \rm MeV$, although not available for the EoSs studied, may provide a better estimation of the thermal EoS.

We begin our discussion by looking at a single, representative EoS for simplicity. Fig.~\ref{fig:binding-energy} shows the nuclear thermal index, Eq.~\eqref{eq:th_nuc},
as a function of density for the HS(IUF) equation of state. The solid lines show the nuclear thermal index as given from the CompOSE dataset for temperatures $T=5$, $15$, $25$ and $45$ MeV. 
We find that the nuclear thermal index at very low densities
is relatively flat and independent of density, but not independent of temperature. At the lowest density considered here, the nuclear thermal index ranges from $\Gamma_\text{th}^\text{nuc} \approx 1.4$ at $T=5$ MeV up to $\approx 1.58$ at $45$ MeV. 
This temperature dependence differs significantly from the results obtained, where the low-density thermal index is essentially density- and temperature-independent, and very close to $5/3$ (see Fig.~\ref{fig:Gamma_neutron}).  It is also differs from the behavior of the nuclear thermal index in $npe\gamma$ matter shown in Fig.~\ref{fig:Gamma_contributions}. We speculate that this effect is largely due to clustering. 
When we compare data slices at different temperatures to compute the thermal pressure and energy density, we are effectively examining mixtures of varying compositions. At low densities and very low temperatures, matter is expected to be in a clustered phase with a wide range of isotopes~\citep{Shen2011,Horowitz06_3}. At higher temperatures ($\gtrsim 15$ MeV), clusters are instead dissolved and matter is homogeneous. 
In order for clusters to dissolve as temperature increases, they must absorb an energy from the medium which is roughly equal to their binding energy. This reduces the available thermal energy and can significantly change the thermal index~\citep{Raduta_2021}. 

\begin{figure}
\includegraphics[width=\linewidth]{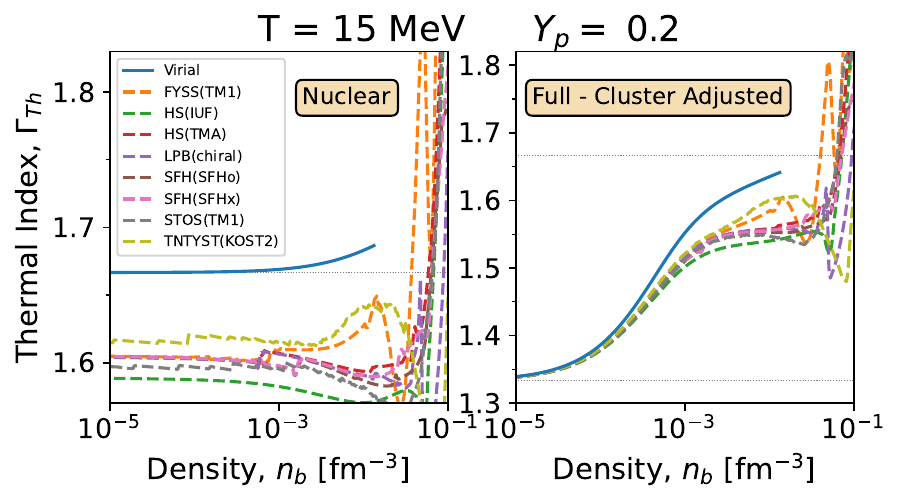}
\includegraphics[width=\linewidth]{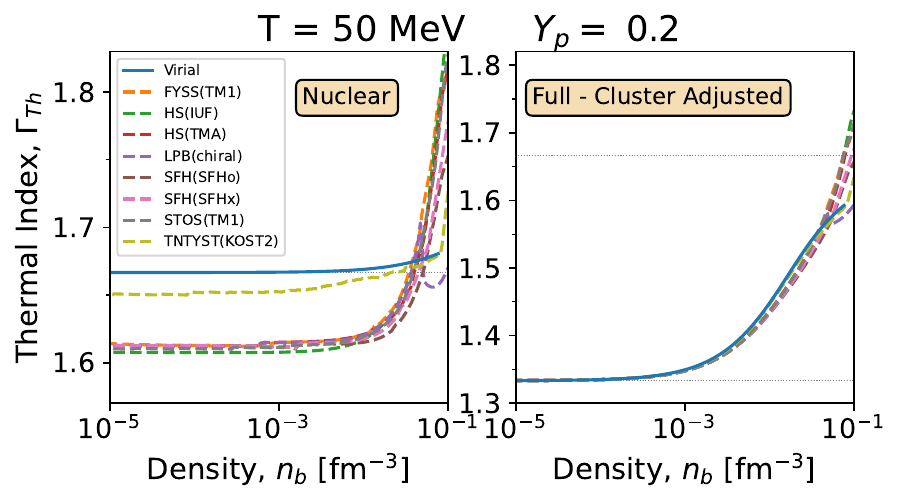}
\caption{\label{fig:comparison}
Thermal index $\Gamma_{\rm th}$ as a function of number density for our virial approximation (blue solid line) and several EoSs from CompOSE (dashed lines).  
The left and right panels display the nuclear thermal index (Eq.~\eqref{eq:th_nuc}) and the full thermal index (Eq.~\eqref{eq:th_index_asymmetric}), respectively,
for a  proton fraction $Y_{p} = 0.2$ and at a temperature of~$T=15$ MeV (top) and at a temperature of~$T=50$ MeV (bottom).
Note the different scales in the $y-$axis of left and right panels. Horizontal dotted lines represent
the values of $\Gamma_\text{th}$ in both the relativistic and non-relativistic limits.} 
\end{figure}

To account for this effect, we add back into the system the energy required to disintegrate nuclei. 
The HS(IUF) CompOSE dataset includes information on cluster composition. We employ this information and compute the binding energy for each species using tabulated data from AME2020 \citep{Huang_2021,Wang_2021} when available. We rely on the liquid drop model otherwise~\citep{Moller:2015fba}. This approach is of course only approximate: it only considers the average cluster given in the CompOSE database (instead of the full nuclear composition) and it does not take into account medium-induced binding energy changes \citep{Typel2010}. 
Introducing these binding energy corrections to remove the effect of clusters, however, changes  the low-density behavior of the nuclear thermal index. We 
display in Fig.~\ref{fig:binding-energy} these cluster-adjusted nuclear thermal indices in dashed lines. 
We observe that removing the effects of clusters approximately restores the temperature independence of the nuclear thermal index for temperatures 
$T>15$ MeV. At low densities, $\Gamma_\text{th}^\text{nuc}$  settles at $\approx  1.6$ for $T$ ranging from $15$ to $45$ MeV.

This approximate cluster adjustment procedure has a similar effect in the nuclear thermal index of a wide variety of EoS.
We show in the left panels of Fig.~\ref{fig:comparison} the nuclear thermal index for the virial EoS (solid blue line) and several database EoSs (dashed lines) at temperatures $T=15$ (top panels) and $50$ MeV (bottom panels). For simplicity, we limit our focus to a fixed proton fraction of $Y_p=0.2$. 
The left panels display the nuclear thermal index (Eq.~\eqref{eq:th_nuc}), while the right panels present the complete thermal index 
(Eq.~\eqref{eq:th_index_asymmetric}), which includes contributions from leptons and photons.
We discuss the nuclear-only index first, before focusing on the full thermal index. 

The virial nuclear thermal index at low densities begins from a value very close to $5/3$ (marked by a dotted line), characteristic of a non-interacting gas. As the density increases, interaction effects become significant and the nuclear thermal index increases to values closer to $\approx 1.7$ before the virial expansion breaks down.
At low temperatures ($T=15$ MeV, top panel), the thermal index of database EoSs show a richer density dependence even after they have been corrected for clustering effects. 
In the density range $n_b \approx 10^{-5}-10^{-3}$ fm$^{-3}$, the nuclear thermal index is roughly constant, with values close to $\Gamma^\text{nuc}_\text{th} \approx 1.6$ for all tabulated EoSs. We take this as an indication that our treatment correctly subtracts the effect of clusters and yields a nuclear thermal index that is close to the homogeneous case. We find a relative sensitivity to the EoS, presumably due to the different nuclear physics ingredients in the selected CompOSE data which our cluster adjustment procedure misses. As density increases, the thermal index generally decreases slightly around $n_b \approx 0.01$ fm$^{-3}$, before showing a rapid increase as density reaches nuclear values close to $n_b \approx 0.1$ fm$^{-3}$. This increase is very sensitive to the EoS model and is likely due to the transition between the NS crust and the core.

The nuclear thermal index in the lower left panel corresponds to results at a higher temperature, $T=50$ MeV, where one would naively expect the clustering to be irrelevant. We observe a relatively density-independent nuclear thermal index at low densities,
$\Gamma^\text{nuc}_\text{th} \approx 1.63$, but this time cluster-adjusted results are much more model independent.
The virial approximation predicts a nuclear thermal index about $\approx 0.05$ larger than the CompOSE data.  
Despite these differences, the discrepancy between the nuclear thermal index of $npe\gamma$ matter at $\beta$-equilibrium within the virial approximation and the cluster-corrected database EoSs never exceeds $\approx 5\%$. 

The discrepancies in the nuclear thermal index at low densities appear to have minimal impact on the full thermal index, as shown in the right panels 
of Fig.~\ref{fig:comparison}. The virial prediction overlaps with the full CompOSE thermal indices across a wide range of densities, up to $n_b \approx 10^{-4}$ fm$^{-3}$ at $15$ MeV (top right panel), and almost up to $0.1$ fm$^{-3}$ at $T=50$ MeV (bottom right panel). The reason is that the region where the nuclear differences are most important is precisely where the nucleonic contributions to thermal effects are less dominant. At low densities, leptons and photons dominate and the overall behavior of a smooth transition between a lepton-dominated at low densities and a nucleon-dominated regime at higher densities is maintained. 
However, once the models reach a certain density, they consistently start to diverge. We take this as an indication of the onset of excluded-volume and nuclear effects.

\begin{figure} 
\includegraphics[width=\linewidth]{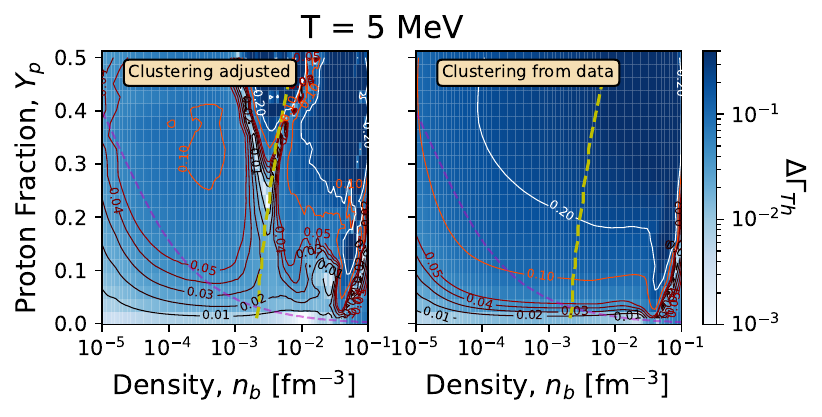}
\includegraphics[width=\linewidth]{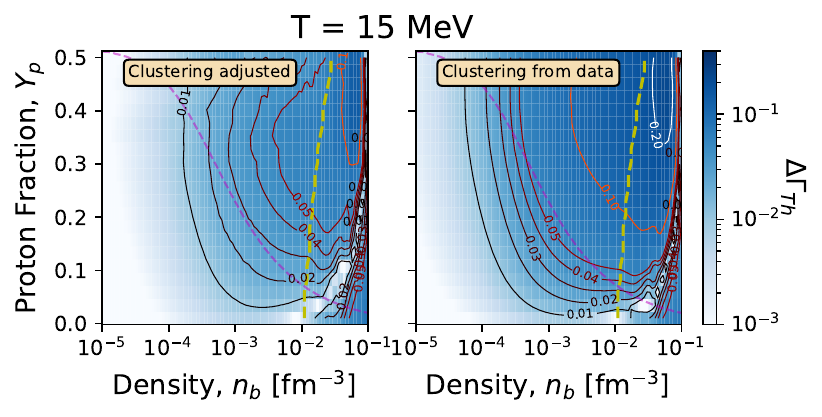}
\includegraphics[width=\linewidth]{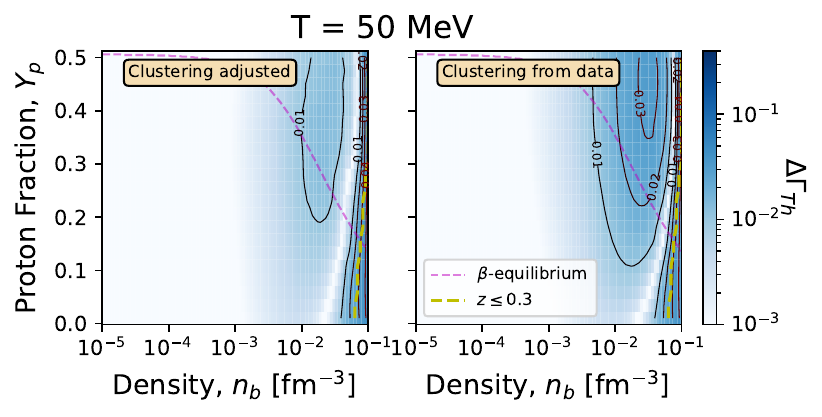}
\caption{
\label{fig:colorplor_comparisond}
Relative error of the virial thermal index compared to the average thermal index from database EoSs for temperatures 
$T=5$ (top), $T=15$ (central) and $50$ MeV (bottom panels). The $y-$axis shows the proton fraction $Y_p$, 
while the 
$x-$axis shows the baryon number density $n_b$ ranging from 
$10^{-5}$  to $10^{-1}\,\text{fm}^{-3}$.  
We show contour lines to guide the eye.
The violet dashed line represents the $\beta$-equilibrium path, while the yellow line indicates the constraint $z_n,z_p\leq0.3$,
as shown in Fig.~\ref{fig:panel}. Each panel depicts the complete EoS, incorporating contributions from leptons and photons. 
In the right panels, the virial expansion is directly compared to the database EoS, whereas in the left panels, the data are 
cluster-adjusted as described in the main text.
} 
\end{figure}

The comparisons so far indicate that the full thermal index of a wide 
variety of CompOSE data is well described by the relatively simple
virial approximation, especially if we remove clustering effects from 
the tabulated data. We provide a broader perspective on these comparisons with the color density plots of Fig.~\ref{fig:colorplor_comparisond}. Here, we display the relative error of our virial approximation, accounting also for leptons and photons, with respect to the average CompOSE behavior, that is 

\begin{align}
\label{eq:relative_error_colorplot}
    \Delta\Gamma_{\rm th}=\frac {\left|\Gamma_{\rm Virial}-\langle\Gamma_{\rm Database}\rangle\right|}{\langle\Gamma_{\rm Database}\rangle},
\end{align}  
where we take the average
\begin{align}
 \langle\Gamma_{\rm Database}\rangle= \frac{1}N_{\rm EoS} \sum_{\rm EoS=1}^{N_{\rm EoS}} {\Gamma_{\rm th,\,EoS}}\,,
\end{align}
over the $N_{\rm EoS}=8$ EoS employed in this comparison. 
The panels provide the density and proton fraction dependence of the relative error for the three representative temperatures $T=5$ (top), $15$ (middle) and $50$ MeV (bottom). We provide indicative contour plots to guide the eye. The dashed lines are the same as those in Fig.~\ref{fig:panel}, indicating the $\beta-$equilibrium trajectory (purple line) and the limit to the virial approximation $z_n,z_p=0.3$ (yellow, bolder line). 
Left panels show results where clusters are removed employing our relatively simple prescription, while right panels correspond to the relative error with respect to the raw CompOSE data. This distinction provides a practical quantification of clustering in the thermal index.

We find a substantial temperature dependence, going from top to bottom panels in Fig.~\ref{fig:colorplor_comparisond}. At $T=5$ MeV (top panels), for instance, the virial expansion only converges for very low densities and only 
exhibits a good match with respect to the average EoS behavior at low proton fraction. We observe relatively large differences with respect to the raw CompOSE data 
$\gtrsim 20\%$~(right panel). When we adjust for the clusters (left panel), the region $n_b < 10^{-3}$ fm$^{-3}$ is relatively well described by the virial $npe\gamma$ thermal index up to $\approx 10\%$. 
Beyond this density and up to $n_b \approx 10^{-2}$ fm$^{-3}$, the differences pile up substantially, reaching values of over $20 \%$, with higher discrepancies for higher proton fractions. Despite these discrepancies, after correcting the data for clusters, and in the region where the virial approximation is valid, the $\beta-$equilibrium thermal index in the virial approximation is within $5 \%$ of the average, clustered-adjusted EoSs. In fact, close to PNM, for proton fractions below $Y_p \approx 0.05$, the relative difference is within $1 \%$. 

The results of the top right panel indicate that, at $T=5$ MeV, the effects of clustering are quite relevant. 
We observe that, without removing the cluster contribution, the virial prediction becomes less accurate across
most densities and proton fractions.
Differences of more than $10 \%$ are consistently observed for proton fractions $Y_p > 0.1$. At this low temperature, the virial approximation only reproduces the CompOSE data for proton fractions $Y_p \lesssim 0.03$.

The relatively large discrepancies between the virial thermal index and the CompOSE averages decrease steeply with temperature. The central panels at $T=15$ MeV illustrate this fact. We observe that, when the EoSs are adjusted for clusterig effects (left panels), the region where $ \Delta\Gamma_{\rm th}$ exceeds $10\%$ is already beyond the validity of the virial approximation. Similarly as in the $T=5$ MeV case, the cluster adjustment is such that the $\beta-$equilibrium thermal index remains within $1-3 \%$ of the average CompOSE result. The raw data (right panel) differs from the virial approximation in a wider range of densities and proton fractions. However,the most proton-rich cases, the virial approximation is within $10-15 \%$ of the average CompOSE result. For proton fractions below $Y_p=0.1$, the virial approximation remains within $5 \%$ of the full database average. 

Finally, the bottom panels show the results obtained at $T=50$ MeV. At this temperature we do not expect clustering to be significant. We indeed observe
that the left and the right panels provide very similar results.  Regardless of clustering processes, we observe that the virial approximation is within a few percent of results provided by CompOSE EoSs. We take this as an indication that, for large temperatures, our approximation provides a valid description of the thermal properties of dense matter, in agreement with more microscopic nuclear models. This may be a consequence of the fact that leptons dominate the thermal properties of matter in this high-temperature regime, to the point that nuclear effects become largely irrelevant. 

To summarize, our findings suggest that the approximation of  $npe\gamma$ matter matter, using the virial expansion 
to account for nuclear effects, closely aligns with the average result predicted by state-of-the-art microscopic tabulated EoS.
Clustering is largely responsible for the differences with respect to tabulated EoSs. Our approach, however, leads to errors $\lesssim 20 \%$ at very low temperatures ($T=5$ MeV), and mostly in the regions where the proton fraction is larger. Along the $\beta-$equilibrium trajectories, for instance, the differences with respect to tabulated EoS is generally smaller within the complete temperature regime. 
A potential assessment of the importance of these discrepancies in numerical simulations could provide further insight on the importance of clustering in BNS physics.


 \section{Conclusion}\label{sec:conclusion}
The thermal properties of the dense matter in the interior of neutron stars play a key role in 
determining the dynamics of binary neutron star systems, influencing phenomena such as tidal 
interactions, gravitational wave emission, and the equation of state.
We provided a description of the thermal index employing the virial expansion to describe nuclear correlations. 

The virial expansion is a model-independent approach to evaluate the thermodynamical properties of an interacting fermionic gas. 
Our approach is reliable at very low densities, starting from a fiducial value of $10^{-5}$ fm$^{-3}$.
The high-density limit of the virial expansion is temperature-dependent, and ranges from about $2 \times 10^{-3}$ fm$^{-3}$ at $T=5$ MeV up to $0.1$ fm$^{-3}$ at $T=50$ MeV,  covering a broad range of densities relevant for binary neutron star mergers.

For neutron matter, the virial expansion suggests a negligible impact of interactions on the thermal index. We observed that, where the virial expansion converges, the thermal index is within a fraction of a percent of the free Fermi gas value $\Gamma_\text{th}=5/3$, independently of density and temperature. Our estimate for truncation errors is an order of magnitude smaller. 
We generalized our findings
to the case of $npe\gamma$ matter employing a well-known extension of the virial expansion to an arbitrary proton fraction, including the effect of deuterons.  The contribution of leptons and photons is always relativistic and dominates at the lowest densities.
As the density increases towards saturation, the thermal index for $npe\gamma$ matter can be understood in terms of a smooth transition from a lepton dominated regime at low densities, with thermal index  $\Gamma_\text{th} \approx 4/3$, to a fermion-dominated region, with  $\Gamma_\text{th} \approx 5/3$, as density increases.
The thermal index can be approximated by a very simple interpolation between these two regimes, Eq.~\eqref{eq:approx}, with an inflection point that depends steeply on temperature, Eq.~\eqref{eq:infpoint}. The effect of nuclear interactions in the density regime of interest is extremely small, within a fraction of a percent.

A limitation of our approach is that it does not account for nuclear clustering.
At the density and temperature regimes of interest for BNS remnants~(see e.g.~\cite{Ruiz:2021qmm,Bamber:2024kfb}, it can modify substantially the thermal properties of matter due to the changing compositions with temperature and the energy absorption that results in the melting of nuclear clusters. Our results indicate that the effect of clustering is relevant for the thermal index at relatively low densities ($T=5$ MeV), where on average microscopic nuclear models may differ up to $\sim 10\%$ from our simple parametrization, particularly in areas of moderate proton fraction, with $Y_p > 0.1-0.2$. However,  Our model can reproduce CompOSE database data at high temperatures ($T=50$ MeV) within a fraction of a percent. More importantly, along the $\beta-$equilibrium trajectory, our predictions for the thermal index are reliable compared to the average CompOSE behavior within 
$10 \%$ ($2 \%$) at temperature of $T=5$ ($T=50$) MeV.

We emphasize that our straightforward approach, based on the virial expansion, closely reproduces the behavior of some tabulated 
EoSs (see Sec.~\ref{sec:comparison}).
The dominant contribution in the evolution of the thermal index at low densities is the transition from the lepton to the nucleon-dominated regimes. Clustering effects are relevant at low densities and temperatures, whereas nuclear interactions are largely irrelevant below saturation density. 
Our numerical results are available in~\citep{Rivieccio_Virial-EoS_2025}.

Our work may be expanded in several ways. 
First, it will be of interest to determine whether the generic transition for the thermal index predicted in our work has an impact on numerical simulations of BNS mergers in the determination of the ejecta properties. While the high-density regime, dominated by nuclear interactions, is interesting in terms of the GW signal, the phase space of BNS simulations is much wider and the observable consequences of such a transition may be relevant. Our work suggests that this effect may have been present in previous simulations employing tabulated data, but a more thorough quantification is required. The simple parametrization proposed in Eq.~\eqref{eq:approx} and Eq.~\eqref{eq:infpoint} may provide a good starting point in this direction.

Our predictions are limited by the range of validity of the virial expansion, which at low temperatures is a substantial constrain. Expanding these towards higher densities is also an important undertaking. We want to extend the virial expansion beyond the second order since our current estimates for the truncation error of the expansion are relatively heuristic~\citep{Hou_2020}. A deeper understanding of the thermal index in the nuclear-density regime may also require the use of more complex nuclear many-body schemes~\citep{Rios2020,Kochankovski2022,Keller2023}.

%
\acknowledgments
\section*{Acknowledgments}
We acknowledge Adriana Raduta, Joaquín Drut and Hristijan Kochankovski for useful 
discussions. This work is supported  by the Agencia~Estatal 
de Investigaci\'on MCIN/AEI/10.13039/-501100011033
(FSE “El FSE invierte en tu futuro” and ERDF A way of making Europe)
through grants PID2020-118758GB-I00, PID2021-125485NB-C21, PID2021-125485NB-C22 and PID2023-147112NB-C22;
through the ``Ram\'on y Cajal" grant RYC2018-026072; and
through the ``Unit of Excellence Mar\'ia de Maeztu 2020-2023" award to the Institute 
of Cosmos Sciences, grant CEX2019-000918-M.  Additional support is provided by the 
Generalitat de Catalunya (AGAUR) through grant  SGR-2021-01069 and
by the Generalitat Valenciana through grants CIDEGENT/2021/046 and Prometeo CIPROM/2022/49. 
%

\appendix
\section{ Numerical values of the virial coefficients}
\label{sec:appendixA}

\begin{table}[t]
\centering
\begin{tabular}{cccccccc}
T [MeV] & $b^{(2)}_n$ (with CIB) & $\overline{ b_n^{(2)}} \equiv \Gamma^{(1)}$ & $\Gamma^{(2)}$ & $b^{(2)}_{np}$ & $\overline{ b^{(2)}_{np} }$ & $b^{(2)}_\text{nuc}$ & $\overline{ b^{(2)}_\text{nuc} }$  \\ \hline
1.0 	&0.2874	& 0.0321	  &0.0099   &18.1309	&-43.4614	&18.4182   &-43.4293	  \\
	2.0 	&0.3023	& 0.0119  &0.0037  	&4.9615	    &-7.2588 	&5.2638	   &-7.2469	      \\
	3.0 	&0.3054	& 0.0037  &0.0011  	&2.9232	    &-3.4110    &3.2285	   &-3.4073	      \\
	4.0 	&0.3059	& 0.0005  &0.0002  	&2.1368	    &-2.1762    &2.4427	   &-2.1757	      \\
	5.0 	&0.3059	& -0.0001 &0.0000  	&1.7202	    &-1.5969    &2.0262	   &-1.5970	      \\
	6.0 	&0.3060	& 0.0007  &0.0002  	&1.4607	    &-1.2673    &1.7666	   &-1.2667	      \\
	7.0 	&0.3062	& 0.0022  &0.0007  	&1.2822	    &-1.0567    &1.5884	   &-1.0545	      \\
	8.0 	&0.3066	& 0.0041  &0.0013  	&1.1512	    &-0.9113 	&1.4577	   &-0.9072	      \\
	9.0 	&0.3072	& 0.0061  &0.0019  	&1.0503	    &-0.8052 	&1.3574	   &-0.7991	      \\
	10.0	&0.3079	& 0.0082  &0.0026  	&0.9698	    &-0.7245 	&1.2777	   &-0.7163	      \\
	12.0	&0.3098	& 0.0121  &0.0038  	&0.8485	    &-0.6103 	&1.1583	   &-0.5982	      \\
	14.0	&0.3119	& 0.0153  &0.0049  	&0.7606	    &-0.5336 	&1.0725	   &-0.5183	      \\
	16.0	&0.3141	& 0.0178  &0.0058  	&0.6931	    &-0.4786 	&1.0072	   &-0.4609	      \\
	18.0	&0.3163	& 0.0194  &0.0064  	&0.6392	    &-0.4375 	&0.9555	   &-0.4181	      \\
	20.0	&0.3184	& 0.0203  &0.0068  	&0.5948	    &-0.4057 	&0.9132	   &-0.3853	      \\
	25.0	&0.3230	& 0.0197  &0.0066  	&0.5107	    &-0.3510 	&0.8337	   &-0.3313	      \\
	30.0	&0.3264	& 0.0157  &0.0053  	&0.4499	    &-0.3167 	&0.7763	   &-0.3009	      \\
	35.0	&0.3287	& 0.0094  &0.0032  	&0.4028	    &-0.2931 	&0.7315	   &-0.2837	      \\
	40.0	&0.3299	& 0.0015  &0.0005  	&0.3647	    &-0.2757 	&0.6946	   &-0.2741	      \\
	45.0	&0.3303	& -0.0073 &-0.0024 	&0.3328	    &-0.2619 	&0.6632	   &-0.2693	      \\
	50.0	&0.3301	& -0.0168 &-0.0054 	&0.3055	    &-0.2504 	&0.6356	   &-0.2672	      
\end{tabular}
\caption{\label{TableAppenA}
Numerical values of the virial coefficients at different temperatures.  We list 
the second virial coefficient $b^{(2)}_n$ for PNM, the first and second order term of the approximation $\Gamma_1$ and $\Gamma_2$ respectively in Eq.~(\ref{eq:th_index_neutrons}), the second virial coefficient for asymmetric matter $b^{(2)}_{np}$, the dimensionless temperature derivative of the latter
$\overline{ b_{np}^{(2)}}=Tb^{(2)\prime}_{np}$, the contribution to the virial coefficient $b^{(2)}_{nuc}$, and its dimensionless derivative temperature $\overline{ b_\text{nuc}^{(2)}}$.
}
\end{table}

Below, we discuss key details relevant to the implementation of the virial expansion. Table~\ref{TableAppenA} displays the numerical values for the relevant virial coefficients for PNM and also for asymmetric matter. Calculations are performed at several temperatures (first column in Table~\ref{TableAppenA}), with the corresponding values of the various coefficients discussed in the main text displayed in the subsequent columns. The integrals for the coefficients (second and fifth columns in Table~\ref{TableAppenA}) are performed using phase shift data from the Granada database for energies up to $350$ MeV~\citep{granada}. Beyond this energy, we assume a constant phase shift, equal to the value $\delta_{\tau \tau'}(E=350 \text{ MeV})$. We have checked that changing this energy threshold has a minimal impact (less than $3 \%$) on the virial coefficients, as contributions from high energies are suppressed by the exponential factors in the integrals. The neutron contribution to the scattering phase shift includes effects from CIB~\citep{Horowitz06_1,Horowitz06_2}. While we employ partial waves up to $J \leq 7$, Ref.~\cite{Horowitz06_1} used $L \leq 6$. Our results for the virial coefficients are however consistent up to $4 \%$ across the whole temperature regime. A publicly available code for our implementation can be found in~\citep{Rivieccio_Virial-EoS_2025}.

In addition to the standard virial coefficients, the calculation of the thermal index requires an accurate calculation of the temperature derivatives $\overline{b_n^{(2)}}$. We do not rely on numerical derivatives with respect to temperature, but rather rewrite the derivatives as
\begin{align}
\overline{ b_n^{(2)}}= T \frac{ d b^{(2)\prime}_{n} }{dT} = 
- \Delta b_n^{(2)} + 
\frac{1}{2 \sqrt{2}\pi T^2} \int^{\infty}_0 \! dE \, E \, e^{-E/2T} \delta^{tot}(E)\,,
\end{align}
and explicitly compute the integral. Once again, we use the QUADPACK package for a fast and reliable calculation of this component~\citep{2020SciPy-NMeth}.

Although the range of validity of the virial expansion primarily applies at high temperatures, we also provide numerical results down to $T=1\,\rm MeV$. We observe that the virial coefficient for neutron matter (second column in Table~\ref{TableAppenA}) increases with temperature up to around $T \approx 41$ MeV. Beyond this point, the virial coefficients
reaches a steady state
and slowly decreases with temperature. In turns, this means that the temperature derivative of the virial coefficient (third column in Table~\ref{TableAppenA}) becomes negative. 
Eqs.~(\ref{eq:th_index_neutrons}) and (\ref{eq:gammas}) suggests that the first-order coefficient for the thermal index, $\Gamma^{(1)}$, depends on the sign of the second-order virial derivative coefficient, $\overline{ b_n^{(2)}}$. This 
has several implications. First, since $\Gamma^{(1)}$ is small, and $\Gamma^{(2)}$ is about a factor of $3$ smaller, the medium modifications to $\Gamma_\text{th}$ are small. Second, when the derivative of the virial coefficient changes sign, the dependence on fugacity of the thermal index also changes. 
In particular, at low temperatures, where $\overline{ b_n^{(2)}}>0$, the thermal index tends to decrease. In contrast, at high temperatures, where $\overline{ b_n^{(2)}}<0$, the thermal index increases with density  (see~Fig.~\ref{fig:Gamma_neutron}).

\section{Virial expansion for a neutron-proton mixture}
\label{sec:appendixB} 
We now turn to discussing the virial expansion in arbitrary isospin-asymmetric matter. This is a well-known approach~\citep{Horowitz06_2}, but we reproduce here the main expressions for completeness. 
 We assume that the pressure is a power series of the neutron $z_n=e^{\mu_n/T}$ and proton $z_p=e^{\mu_p/T}$ fugacities. To second order, this expansion is given by
\begin{align}
    P=\frac{2T}{\lambda^3}\,\left[ 
    z_n+z_p+
\left(z_n^2+z_p^2\right)
    \,b^{(2)}_n+2\,z_n\,z_p\,b^{(2)}_{np}
    \right],
    \label{eq:press_asymmetric}
\end{align}
where $b^{(2)}_{np}$ is the second virial coefficient for the $np$ interaction. We employ the same $b^{(2)}_n$ from Sec.~\ref{section:virial} as the second virial coefficient for both $nn$ and $pp$ interactions.
The thermal de Broglie wavelength $\lambda$ is computed with the average nucleon mass. This approximation neglects Coulomb interaction effects and assumes that nuclear interactions are charge-independent.

The coefficient $b^{(2)}_n$ is obtained as described in Eq.~(\ref{eq:b2_n}), while the coefficient $b^{(2)}_{np}$ can be decomposed into different isospin components,
$b^{(2)}_{np}(T)=b^{(2)}_{nuc}(T)-b^{(2)}_n(T)$.  We explicitly include the deuteron as a bound state contribution to $b^{(2)}_{nuc}$, with experimental binding energy of $E_d=2.22$ MeV~\citep{Horowitz06_2}. This contribution, together with the scattering $np$ phase-shifts, yields 
\begin{align}
b^{(2)}_\text{nuc}(T)&=-2^{-5/2} + \frac{3}{\sqrt{2}}\left(e^{E_d/T}-1\right)
+\frac{1}{2^{3/2}\pi T} \int^{\infty}_0 \! dE \, e^{-E/2T} \delta^{tot}_{nuc}(E),
\label{eq:b2_np}
\end{align}
where $ \delta^{tot}_{nuc}$ is now the sum over all partial waves and includes degeneracy factors depending on the isospin, $I$, and the total angular momentum $J$,
\begin{align}
    \delta^\text{tot}_\text{nuc}(E)&=\sum_{S,L,J}(2J+1)(2I+1)\delta_{^{2S+1}L_J}(E)
    = 3\delta_{^{1}S_0}+3\delta_{^{3}S_1}+3\delta_{^{1}P_1}+3\delta_{^{3}P_0}+\cdots\,.
\end{align}
As for the PNM described in Sec.~\ref{section:virial}, we assume that for energies beyond our data range $\delta_{tot}$ remains constant. 

The virial coefficient for asymmetric matter (fifth column in Table~\ref{TableAppenA}) peaks at low temperatures, driven by the presence of the deuteron bound states which enhances the $b^{(2)}_{nuc}$ coefficient  (seventh column in Table~\ref{TableAppenA}) at low densities. As the temperature increases, however, the deuteron bound state melts, and the $np$ virial coefficients become smaller. The temperature-derivative-coefficients in columns $6$ and $7$ in Table~\ref{TableAppenA} start from large negative values at low temperatures and subsequently become smaller in magnitude. We find that, since 
$\overline{ b_n^{(2)}}$ is very small, the two derivative coefficients are essentially the same, 
$\overline{ b_{np}^{(2)}} \approx 
\overline{ b_\text{nuc}^{(2)}}$.

The densities for neutrons and protons are obtained by differentiating the pressure with respect to the corresponding fugacities, 
\begin{align}
n_{\tau}&=\frac{z_{\tau}}{T}\left(\frac{\partial P}{\partial {z_{\tau}}} \right)_{V, T}
    =\frac{2}{\lambda^3}\left(z_{\tau}+2\,z_{\tau}^2\,b^{(2)}_n+2\,z_n\,z_p\,b^{(2)}_{np}\right),
\end{align}
with $\tau= n,p$.
The entropy density is obtained by differentiating the pressure with respect to temperature, 
\begin{align}
s=\left(\frac{\partial P}{\partial T}\right)_{\mu_n,\mu_p}&=\frac{5P}{2T}-n_n\log z_n - n_p\log z_p +\frac{2}{\lambda^3}\left[ 
(z_n^2+z_p^2)\overline{b_n^{(2)}}+2z_nz_p \overline{b_{np}^{(2)}}
\right]\,.
\end{align}
Finally, the energy density is calculated from the entropy density and the pressure,
\begin{align}
    \epsilon &=  T\,s+\sum_{i=n,p}n_i\mu_i - P
    =
    \frac{3}{2}P 
    +\frac{2T}{\lambda^3} \left[
    (z_n^2+z_p^2)\, \overline{b_n^{(2)}}+2\,z_n\,z_p \overline{b_{np}^{(2)}}
    \right].
\label{eq:ener_asymmetric}
\end{align}
As in Sec.~\ref{Sec:virrial_exp}, by examining Eqs. ~(\ref{eq:press_asymmetric}) and (\ref{eq:ener_asymmetric}) and comparing them to Eq.~~(\ref{eq:press_descomposada}) and Eq.~~(\ref{eq:ener_descomposada}), it is evident that both the pressure and energy of the system are intrinsically thermal and hence $P^{nuc}_{\rm th}=P$ 
and $\epsilon^{nuc}_{\rm th}=\epsilon$. With this, one can readily obtain an expression for the nuclear thermal index up to second order in the virial expansion of the pressure and the energy density, 
\begin{align}
    \Gamma_\text{th}^\text{nuc} &= 
    \frac{5}{3}
    - \frac{4}{9} 
    \frac{
    (z_n-z_p)^2 \overline{b_n^{(2)}} 
    + 2 z_n z_p \overline{b_\text{nuc}^{(2)}} }{
    z_n + z_p + (z_n-z_p)^2 b_n^{(2)}
    + 2 z_n z_p b_\text{nuc}^{(2)}
    + \frac{2}{3} 
    \left[ (z_n-z_p)^2 \overline{b_n^{(2)}} 
    + 2 z_n z_p \overline{b_\text{nuc}^{(2)}} \right] 
    }
    \, .
\label{eq:th_asymmetric}
\end{align}
In our calculations, we use this expression computing 
$\Gamma_\text{th}^\text{nuc}$
explicitly from the ratio of thermal energies and pressures. We note that, just like in PNM, the leading virial expansion modifications to the nuclear thermal index arise from the temperature derivatives of the virial coefficients, 
$\overline{b_n^{(2)}}$ and $\overline{b_\text{nuc}^{(2)}}$, rather than the coefficients themselves. 

\bibliography{biblio}
\end{document}